\documentclass{article}

\usepackage{amsmath,amssymb}
\usepackage{lmodern}
\usepackage{iftex}
\usepackage{graphicx}  
\usepackage{amsmath}   
\usepackage{amssymb}   
\usepackage{bm} 
\usepackage{dcolumn}
\usepackage{color}
\usepackage{mathrsfs}
\usepackage{amsfonts}
\usepackage{varioref}
\usepackage[margin=1.4in]{geometry}
\usepackage{slashed}
\usepackage{authblk}
\usepackage{float}
\usepackage{subcaption}
\usepackage{mathtools}
\usepackage{pst-node}
\usepackage{tikz}
\usetikzlibrary{tikzmark}
\restylefloat{figure}
\RequirePackage[colorlinks,citecolor=blue,urlcolor=magenta,linkcolor=blue]{hyperref}
\def\p{\partial}
\def\e{\epsilon}

\def\be{\begin{equation}}
\def\ee{\end{equation}}

\labelformat{section}{Section #1} 
\labelformat{subsection}{Section #1} 
\labelformat{subsubsection}{Section #1}
\labelformat{subsubsubsection}{Section #1}
\labelformat{equation}{Eq.~(#1)} 
\labelformat{figure}{Fig.~#1} 
\labelformat{subfigure}{Fig.~\thefigure#1} 
\labelformat{table}{Tab.~#1} 
\labelformat{appendix}{Appendix #1}
\title{\bf Spontaneous symmetry breaking induced by curvature  :  Analysis via non-perturbative 2PI Hartree approximation}
\author[]{Vishal Nath\footnote{vishaln.physics.rs@jadavpuruniversity.in} ,   \, Kinsuk Roy\footnote{kinsukr.physics.rs@jadavpuruniversity.in} \, and   Sourav Bhattacharya\footnote{sbhatta.physics@jadavpuruniversity.in}}
\affil[]{Relativity and Cosmology Research Centre, Department of Physics, Jadavpur University,\\ Kolkata 700 032, India. }

\begin{document}
\maketitle
\begin{abstract}
\noindent
In this work we investigate the spontaneous symmetry breaking (SSB) induced by a classical background spacetime's curvature, via the 2 particle irreducible (2PI) non-perturbative effective action formalism. We use  the standard Schwinger-DeWitt local expansion of the Feynman propagator, appropriate to probe the effect of spacetime curvature on the local or short scale physics.  Recently it was shown using perturbative computations that such SSB is possible with a scalar with a quartic self interaction,  positive rest mass squared and positive non-minimal coupling. Here we confirm in the two loop Hartree approximation  that curvature can indeed induce SSB for such a theory. SSB  for such a model is  not possible in a flat spacetime. The 2PI technique does not only resum  the self energy resulting in mass generation, but also resums, as we have discussed,  curvature terms through such mass generation.  We have explicitly discussed our results  in the context of the de Sitter spacetime, although our calculations are valid for any non-singular curved spacetime. We show that, in contrast to the perturbative results, SSB is possible with a vanishing non-minimal coupling.  These results are further extended to the case of an $O(N)$ symmetric scalar field theory. Restoration of the broken symmetry in the thermal case is also briefly discussed. 
\end{abstract}
\vskip .5cm


\pagebreak
\tableofcontents
\section{Introduction}\label{S1}
Quantum field theory in curved spacetime is considered  as a bridge between the standard quantum field theory and hitherto unknown theory of quantum gravity. This topic is old, and has provided us various theoretical breakthroughs like the Hawking radiation, Unruh effect, stress energy tensor renormalisation, loop correction of the cosmological correlation functions  and so on. We refer our reader to~\cite{Dewitt, Birrell:1982ix, Wald:1995yp, Toms, Senatore:2009cf, Tanaka:2013caa} and references therein for a vast review and list of references on these topics. In this work we will be interested in the spontaneous symmetry breaking and its restoration  in curved spacetimes. 

 The spontaneous symmetry breaking mechanism via a scalar potential is a very important phenomenon in standard quantum field theory~\cite{Goldstone:1962es, Coleman} (see also \cite{Espinosa:1992gq} and references therein). This mechanism can give mass to initial massless fields. At high temperature such broken symmetry can be restored, e.g.~\cite{dolan, Wein, Barry, Rod, Paul, Tanguy, Pole, Zel, Ashok} and references therein. The symmetry breaking effect can be generated via an effective potential as an outcome of  radiative corrections, see~\cite{Coleman} for the original seminal work. (see also e.g.~\cite{Jackiw, Ford, Arafune:1997kp, Inagaki:1997gi, Bordag:2002sf, herr} and references therein for various aspects of effective potential, both   at zero and finite temperatures). The community has given efforts to understand the  effect of spacetime curvature on effective action/potential, see e.g.~\cite{Toms, Ishi, Odintsov, Buch, Buchbinder:1992rb, Elizalde:1993ee, dtoms, Toms:2019erd}. We  refer our reader to  e.g~\cite{Birrell:1982ix, Bunch:1980bs,  bunch, Panangaden:1981qj, Parker, Shore} and references therein for discussion on renormalisation, effective action and various other aspects of quantum field theory in general curved spacetimes. We further refer our reader to e.g.~\cite{Prokopec:2011ms, Arai:2012sh, Arai:2013jna,  LopezNacir:2013alw, LopezNacir:2014fhi, LopezNacir:2016gzi, LopezNacir:2020dik, Serreau:2011fu, Gautier:2015pca, Guilleux:2016oqv, Miao:2020zeh, Eriksson:2023lde, Akhmedov:2013vka, Akhmedov:2024npw, Bhattacharya:2025jtp} for discussion on various non-perturbative quantum effects, effective action or effective potential including that of the $O(N)$ symmetric scalar field theory (e.g.~\cite{Eli, Romatschke:2023ztk, Romatschke:2024hpb}, and references therein) in the de Sitter spacetime background. We further refer our reader to~\cite{blhu, Hu:1982ue, naka, Hu, dow,Cog,DeNardo:1997gn, Hat, Hay, Buchholz:2006iv, Kalinichenko, Bhattacharjee:2012my, kwan, Chang:2019ebx, Khakimov:2023emy, Nath:2024doz} and references therein for various finite temperature effects, including that of the effective action or potential in curved spacetimes like the de Sitter. \\

\noindent
It is well known that in the flat or Minkowski spacetime, the simplest way to obtain a spontaneous symmetry breaking (SSB) would be to consider a potential
$$V(\phi)= -\frac12 m_0^2 \phi^2 +\frac{\lambda \phi^4}{4!}\qquad  (m_0^2 {\ensuremath >} 0)$$
The negative mass term creates a maximum at $\phi=0$, two minima and hence a Mexican hat type potential. It was shown in  \cite{Coleman} that the effect of the negative rest mass squared can be generated via loop effects or the effective action, for $m_0^2=0$. To the best of our knowledge,   there can be  no SSB for  $m_0^2 {\ensuremath >} 0$. Let us consider a curved spacetime now.  Can the spacetime curvature generate SSB, which is otherwise not possible in a flat spacetime? For example, can curvature dependent radiative corrections generate a symmetry breaking effective potential with a positive rest mass squared? We wish to address this issue in this paper. In \cite{Shore}, SSB was investigated by modifying the rest mass squired via $R \phi^2$ non-minimal coupling. In~\cite{Hay}, SSB was studied in spacetimes with non-trivial topologies and was shown that a positive (negative) cosmological constant decreases (increases) SSB. 

Recently in~\cite{Nath:2024doz}, the Schwinger-DeWitt  local expansion  for the Feynman propagator in general curved spacetimes~\cite{Dewitt, Toms} was used to compute the effective action for a $\phi^4$-theory at one and two loop and it was shown that SSB is indeed possible for the effective potential, with positive rest mass squared {\it and} a positive non-minimal coupling, in the de Sitter spacetime. The expansion of the propagator up to the quadratic order  in	 the curvature  was used. In particular, no SSB was found in the linear order in the curvature. The above is an example of pure curvature driven SSB. Now, it is very legitimate to ask, what happens if we increase the order of the curvature expansion? Does the SSB feature go away, or it gets more intense? Motivated by this, we wish to address this problem using the  non-perturbative 2 particle irreducible  (2PI) effective action formalism (see~\cite{Berges} for a vast review) in this paper. We will restrict our computations to two loop Hartree, or local approximation. The 2PI formalism effectively resums the self energies via the Schwinger-Dyson equation. This will eventually lead to, as we will see in the due course, resummation of the curvature terms of the propagator as well.  To the best of our knowledge, such curvature driven SSB phenomenon might be physically important, in particular, in the early universe scenario. 

We also note that the propagator one uses in the Schwinger-DeWitt technique is constructed from the local Lorentz invariance and hence is essentially meant to probe the effect of the spacetime curvature on short scale, high energy or ultraviolet quantum processes, such as the trace anomaly~\cite{Dewitt}, or in the processes associated with quantum fields having a compact support. Such propagators may not be appropriate  to understand the non-local, infrared or large scale physics. We refer our reader to~\cite{Prokopec:2011ms,  Arai:2012sh, Arai:2013jna, LopezNacir:2013alw, LopezNacir:2014fhi, LopezNacir:2016gzi, Serreau:2011fu, Bhattacharya:2025jtp} and references therein for computations of the effective potential in de Sitter using the exact propagator. \\

\noindent
The rest of the paper is organised as follows. In the next section, we briefly review the basic set up we will be working in. In \ref{S3} we compute the effective potential using the non-perturbative  2PI Hartree approximation for $\phi^4$ theory with positive rest mass squared. We confirm that SSB is present. Comparison of our result with the perturbative computation of~\cite{Nath:2024doz} has been pointed out. In \ref{S5} we demonstrate the restoration of broken symmetry at high temperatures.  Next in \ref{ON0}, we discuss the same issues for the $O(N)$ symmetric quartic scalar field theory, both in the symmetric and broken phases. Finally, we conclude in \ref{concl}. We will work with the mostly positive signature of the metric in $d=4-\e$ ($\e=0^+$) dimensions, and will set $c=\hbar = 16\pi G=k_B=1$ throughout. We will take the spacetime to be purely classical background  and will ignore any quantum gravity fluctuations. 

\section{The basic set up}\label{Set}

We wish to first briefly review below the basic technical set up we will be working in. It has two parts. The first is the construction of a {\it local} momentum space representation of curvature expansion of the Feynman propagator for a scalar field in a general non-singular  spacetime. We will restrict the expansion up to the quadratic order in the curvature. The second is a very brief review on the 2PI formalism and the Hartree or local approximation scheme for a scalar field theory with quartic self interaction.

\subsection{Feynman propagator in a general curved spacetime}\label{S2'}
\noindent
Let us first discuss the issue of the Feynman propagator. We will work with the Schwinger-DeWitt expansion of the same in a neighbourhood covered by a Riemann normal coordinate system~\cite{Dewitt, Toms, Parker} (also references therein). The action for the scalar field reads
\begin{eqnarray}
S = -\int d^d x\sqrt{-g} \left[\frac12 g^{\mu\nu}(\nabla_{\mu} \phi)(\nabla_{\nu} \phi) +\frac12 m_0^2 \phi^2 +\frac{\lambda\phi^4}{4!} + \frac12\xi R \phi^2\right], \qquad \qquad (\lambda {\ensuremath >} 0)
\label{w1}
\end{eqnarray}
with the corresponding equation of motion,
\begin{eqnarray}
\left(-\square +m_0^2 +\frac{\lambda\phi^2}{3!} +\xi R\right)\phi = 0.
\label{w2}
\end{eqnarray}
As we have stated earlier, the Schwinger-DeWitt expansion of the propagator is essentially meant to capture the effect of the background spacetime curvature in  short scale, local or high energy quantum processes. Accordingly as a suitable frame to probe such phenomena,  one erects a Riemann normal coordinate system in the neighbourhood of a given spacetime point, 
\begin{eqnarray}
&& g_{\mu\nu} = \eta_{\mu\nu} -\frac13 R_{\mu\alpha\nu\beta} y^{\alpha} y^{\beta} - \frac16 R_{\mu\alpha\nu\beta;\gamma} y^{\alpha} y^{\beta}y^{\gamma} +\left(-\frac{1}{20}R_{\mu\alpha\nu\beta;\gamma\delta} + \frac{2}{45}R_{\alpha\mu\beta\lambda}R^{\lambda}{}_{\gamma\nu\delta}\right)y^{\alpha} y^{\beta}y^{\gamma}y^{\delta}\cdots , \nonumber\\
&& |g| = 1- \frac13 R_{\alpha\beta}y^\alpha y^\beta -\frac16 R_{\alpha\beta;\gamma}y^{\alpha} y^{\beta}y^{\gamma} +\left(\frac{1}{18} R_{\alpha\beta} R_{\gamma\delta} -\frac{1}{90} R^{\lambda}{}_{\alpha\beta}\,^{\rho} R_{\lambda\gamma\delta\rho} - \frac{1}{20} R_{\alpha\beta ; \gamma\delta}  \right)y^{\alpha} y^{\beta}y^{\gamma}y^{\delta}\cdots, 
\label{w4}
\end{eqnarray}
where the curvature terms appearing above serve as expansion coefficients, evaluated at the origin of the normal coordinate system, $y=0$. Thus \ref{w4} can be thought of as an expansion analogous to that of the Taylor series.  It is in an explicitly {\it locally} flat form, and  the expansion holds good only if we are interested in a length scale small compared to the characteristic  scale associated with the spacetime $\sim R^{-1/2}$, over which the variation of the curvature is effective. The contractions in \ref{w4} are done with respect to the Minkowski metric. Thus any scalar quantity computed based upon the above expansion will be locally Lorentz invariant.  

Let $G(x,x')$ be the tree level Feynman propagator for a scalar field with mass $m^2$,
$$\left(-\square +m^2\right)i G(x,x')=-i\delta^d (x,x'), $$
where $\delta^d(x,x')= \delta^d(x-x')/\sqrt{-g}$ is the covariant  $\delta$-function.
 For computational conveniences, one next defines  $\bar{G}(x,x^{\prime})$ as
\begin{eqnarray}
G(x,x^{\prime}) = g^{-{\frac{1}{4}}}(x)\bar{G}(x,x^{\prime})g^{-{\frac{1}{4}}}(x^{\prime}).
\label{w5}
\end{eqnarray}
We next plug the above decomposition into the equation for the tree level Green function, and make a systematic expansion of the left hand side in the powers of the curvature.  One can next use   the local flatness of the spacetime, \ref{v4}, in order to define a {\it local $d$-momentum space}, leading to a Fourier decomposition of the Green function owing to Bunch and Parker~\cite{Parker},
\begin{eqnarray}
\bar{G}(x,x^{\prime}) = \int \frac{d^d x}{(2\pi)^d}e^{ik\cdot y} \bar{G}(k)
\label{w6}
\end{eqnarray}
By the virtue of the local expansion of \ref{w4}, we  next expand $\bar{G}(k)$ as
\begin{eqnarray}
\bar{G}(k) = \bar{G_0}(k) +\bar{G_1}(k) +\bar{G_2}(k) +\bar{G_3}(k) +\cdots 
\label{w7}
\end{eqnarray}
where $\bar{G_i} (k)$ contains the $i^{\rm th}$ derivative of the metric evaluated at the centre of the Riemann normal coordinates, $y = 0$.\\

\noindent
One finds at the leading order  the usual flat space Feynman propagator  
\begin{eqnarray}
\bar{G_0}(k) =\frac{1}{k^2 +m^2}
\label{w8}
\end{eqnarray}
Since there is no first derivative of the metric surviving at $y = 0$, we must have $\bar{G_1}(k) =0$. Remembering that the curvature terms appearing in \ref{w4} are just expansion coefficients evaluated at the centre of the normal coordinate system,  one also computes
\begin{eqnarray}
\bar{G_2}(k) =\frac{\left(\frac16 -\xi\right)R}{(k^2 +m^2)^2}, \qquad \qquad 
\bar{G_3}(k) =\frac{i\left(\frac16 -\xi\right)(\nabla_{\alpha} R)}{k^2 +m^2} \p^{\alpha}_{k}\frac{1}{k^2 +m^2}, \nonumber\\
\bar{G_4}(k) =\frac{\left(\frac16 -\xi\right)^2 {R^2}}{(k^2 +m^2)^3} +\frac{a_{\alpha\beta}}{k^2 +m^2}\p^{\alpha}_k\p^{\beta}_k\frac{1}{k^2 +m^2},
\label{w9}
\end{eqnarray}
where we have abbreviated 
\begin{eqnarray}
a_{\alpha\beta} =\frac12\left(\xi -\frac16\right)\nabla_\alpha \nabla_\beta R + \frac{1}{120} \nabla_\alpha \nabla_\beta R -\frac{1}{40} \square R_{\alpha\beta} +\frac{1}{30} R_{\alpha}\,^{\lambda} R_{\lambda\beta} -\frac{1}{60} {R^\rho\,_\alpha}^{\lambda}\,_{\beta} R_{\rho\lambda} -\frac{1}{60} R_{\lambda\mu\rho\alpha} R^{\lambda\mu\rho}{}_{\beta}.
\label{w10}
\end{eqnarray}
We will restrict our calculation to the quadratic order of the spacetime curvature. Up to this order we collect all the relevant terms to write
\begin{eqnarray}
\bar{G}(k) =\frac{1}{k^2 +m^2} +\frac{\left(1/6 -\xi\right)R}{(k^2 +m^2)^2} +\frac{i}{2} {\left(\frac16 -\xi\right)(\nabla_\alpha R}){\p^{\alpha}_k}(k^2 +m^2)^{-2} +\frac13 a_{\alpha\beta}\p^{\alpha}_k\p^{\beta}_k(k^2 +m^2)^{-2} \nonumber\\ +\left[\left(\frac16 -\xi\right)^2 {R^2} -\frac23 a^{\lambda}\,_{\lambda}\right]\frac{1}{(k^2 +m^2)^3} +{\cal O}(R^3).
\label{w11}
\end{eqnarray}

\noindent
We note that, as it is evident from the above discussions, the Schwinger-DeWitt expansion of the propagator \ref{w11} is good to make short or UV scale computations only, such as the trace or chiral anomalies~\cite{Toms}. We cannot use it for non-local or IR computations. Accordingly, we will apply it below to calculate the effective action in the Hartree or local approximation.

\subsection{The 2PI effective action in Hartree or local approximation}\label{S2}

Let us now consider a scalar field theory with quartic self interaction, \ref{w1}. We decompose the field into a classical background ($v$), and a quantum fluctuation ($\varphi$) over it,
$$\phi = v+ \varphi $$
The 2PI effective action which is a functional of the background $v$, as well as the  full or non-perturbative Feynman propagator $iG(x,x')$, is given by~\cite{Berges, Berges:2005hc, Kainulainen:2021eki} (also references therein)
\begin{eqnarray}
\Gamma_{\rm 2PI}[v,iG]=S[v]-\frac{i}{2} \ln \mathrm{det} [iG(x,x')]+\frac{1}{2}  {\rm Tr}\int \sqrt{-g}   \sqrt{-g'} d^dx d^d x' (iG_0 [v](x,x'))^{-1} iG(x',x)+i\Gamma_2[v ,iG]
\label{v1}
\end{eqnarray}
where 
\begin{equation}
(i G_0)^{-1}[v](x,x')= \frac{\delta^2 S[v]}{\delta v(x) \delta v(x')}, 
\label{v2}
\end{equation}
is the inverse of the tree level propagator.
$i\Gamma_2[v,G]$ in \ref{v1} generates all the vacuum graphs which are 2PI. 

\begin{figure}[H]
\begin{center}
\includegraphics[scale=.6]{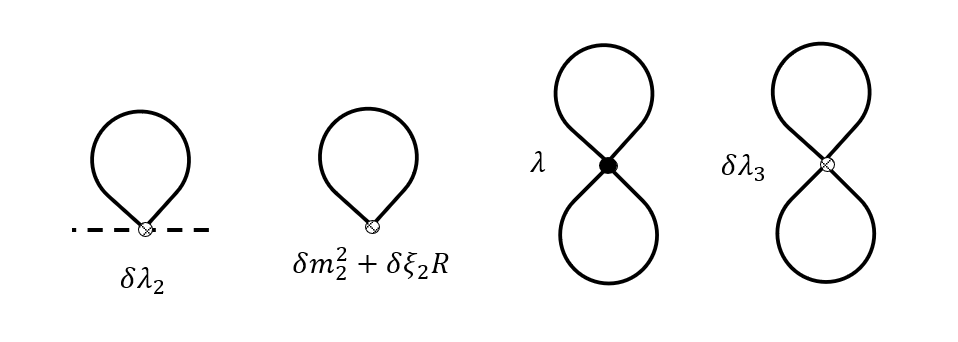}
\caption{  \it \small Various Feynman diagrams for the $\lambda\phi^4$ theory pertaining to the two loop Hartree approximation. The propagators are exact.}
\label{symb100}
\end{center}
\end{figure}

We will restrict ourselves to two loop and {\it local } approximation, with the 2PI vacuum graphs given by the last two of \ref{symb100}, i.e. the double bubble graphs. Note that such graphs, being evaluated at a single spacetime point, make only local contributions.  

\noindent
In this case, the 2PI effective action in a curved spacetime is given by~\cite{Berges},
\begin{eqnarray}
   && \Gamma_{\rm 2PI} [v,iG]=S[v]-\frac{i}{2} \ln \mathrm{det} [iG] 
-\frac{1}{2} \! \int \sqrt{-g} d^d x \left[- \Box +m_0^2+\delta m_{2}^{2} +(\xi+\delta \xi_2)R +\frac{(\lambda +\delta \lambda_2) v^2}{2}  \right] iG(x,x)\nonumber\\&& -\frac{(\lambda +\delta \lambda_3)}{2^3} \int \sqrt{-g} d^d x  (iG(x,x))^2.
\label{v4}
\end{eqnarray}
Note that the last line in the above equation corresponds  to the two loop vacuum graphs of \ref{symb100}. Any field strength renormalisation will not be necessary at two loop. $S[v]$ stands for the classical action corresponding to the background field,
\begin{equation}
S[v]=-\! \int \! \! d^dx \sqrt{-g} \Bigl[ \frac{1}{2} v \left(-\Box +m_0^2+\delta m_{1}^{2}+(\xi  +\delta \xi_1) R\right)v+\frac{(\lambda+\delta \lambda_1) v^4}{4!}  \Bigr]
\label{v5}
\end{equation}
In order to renormalise the effective action, we will also require some gravitational counterterms,
\begin{equation}
\delta S_{\rm grav}= \int \! \! d^d x \sqrt{-g} \left( \delta\kappa R-2 \delta\Lambda+\delta \alpha R^2+\delta \beta R^{\mu \nu} R_{\mu \nu}+\delta \gamma R^{\mu \nu \rho \sigma}R_{\mu \nu \rho \sigma} \right)
\label{v6}
\end{equation}

\noindent
Before we end this section, we note a couple of things which will be more evident from our computations appearing below. 
The variation of \ref{v4} with respect to the Green function yields a Schwinger-Dyson equation. Since the Green function is exact here, this equation is non-perturbative. In the Hartree approximation, the self energy term in this equation is just the one loop  bubble diagram with {\it exact } propagator (originated from the last term on the right hand side of \ref{v4}) and hence is purely local. In other words the Schwinger-Dyson equation  resums those local self energy to yield a dynamical, effective and non-perturbative rest mass of the scalar field. Note that in the standard 1PI effective action technique, where one uses the tree level propagator and there is no such equation for the same,  the effective mass term is simply given by $m_0^2 +\lambda v^2/2$. Here instead we will have the self energy contribution to the mass term  as well.

We will use the propagator \ref{w11} in our calculation of the effective action. 
Now in a curved spacetime,  the spacetime curvature will contribute to field's self energy. Thus while taking \ref{w11} as a non-perturbative propagator in the Hartree approximation, we must take $m^2$ to be the aforementioned dynamical effective mass squared, containing curvature contributions originating through the self energy.  It is thus clear that in this way we not only perform the usual loop resummation associated with the 2PI formalism, but also we resum the curvature terms. Let us now see these things explicitly in what follows.

\section{Renormalisation of the equations of motion in a general curved spacetime}\label{S3}

We now take the variations of the effective action \ref{v4} with respect to $iG(x,x')$ and $v$ to respectively find the following equations of motion,
\begin{equation}
 \left[- \Box_x+m_0^2+\delta m_{2}^{2}+(\xi+\delta \xi_2) R+\frac{(\lambda+\delta \lambda_2) v^2}{2} +\frac{(\lambda+\delta \lambda_3)}{2}  iG(x,x) \right] iG(x,x')
=- \frac{i\delta^d (x-x')}{\sqrt{-g}},
\label{v8}
\end{equation}
and
\begin{equation}
\left[- \Box+m_0^2+\delta m_{1}^{2}+(\xi+\delta \xi_1) R+\frac{(\lambda+\delta \lambda_1) v^2}{6} +\frac{ (\lambda+\delta \lambda_2)}{2} G(x,x) \right] v(x)=0.
\label{v9}
\end{equation}
We have  used above for the inverse tree level propagator, \ref{v2},
\begin{equation}
(i G_0)^{-1}[v](x,x')= \frac{\delta^2 S[v]}{\delta v(x) \delta v(x')}= \left[- \Box_x+m_0^2+\delta m_{2}^{2}+(\xi+\delta \xi_2) R+\frac{(\lambda+\delta \lambda_2) v^2}{2} \right] \delta^d(x,x').
\label{v2'}
\end{equation}
Note in \ref{v8} that $x$ can be taken as the origin of the normal coordinates. Thus on the right hand side when we put $x'=x$ by the virtue of the $\delta$-function, we may put  $\sqrt{-g}=1$, \ref{w4}. Also, each of the above equations are self renormalised. There is no scope nor any necessity to add further counterterms. \\

\noindent
In the Schwinger-Dyson equation \ref{v8}, the $iG(x,x)$ term is the one loop bubble, and it yields a completely local self energy. 
Since the associated propagators are non-perturbative, this self energy is non-perturbative as well. The local self energy will contribute to the rest mass of the scalar field, resulting in an effective, dynamical mass. In other words, we assume that   \ref{v8} can be rewritten in an explicitly ultraviolet finite form as 
\begin{eqnarray}
&& \left[-\Box_x+m_{\mathrm{dyn,eff}}^2+\xi R\right] iG(x,x')=-i \delta^4 (x-x')
\label{v11'}
\end{eqnarray}
Comparing now \ref{v8}, \ref{v11'}, we have for the self consistent renormalisation
\begin{equation}
m_{\rm{dyn,eff}}^2+\xi R=m_0^2+\delta m_2^2+(\xi+\delta \xi_2) R+\frac{(\lambda+\delta \lambda_2) v^2}{2} +\frac{(\lambda+\delta \lambda_3) }{2} G(x,x).
\label{v10}
\end{equation}

We may determine $m_{\mathrm{dyn,eff}}^2$ appearing in \ref{v11'} as follows. 
It is clear that \ref{v11'} will be satisfied by \ref{w11}, provided we replace $m^2$ by $m^2_{\rm dyn, eff}$. In this case we obtain 
\begin{eqnarray}
&&iG(x,x)=\int \frac{d^dk}{(2\pi)^d}\left[\frac{1}{(k^2 +m_{\rm dyn, eff}^{2})} + \frac{\left(1/6 -\xi \right)R}{(k^2 +m_{\rm dyn, eff}^{2})^2} + \frac{2f_1}{(k^2 +m_{\rm dyn, eff}^{2})^3}\right] \nonumber\\&& 
 =-\frac{\mu^{-\epsilon} (m_{\rm{dyn,eff}}^2+(\xi-1/6)R)}{8\pi^2\e}+\frac{m_{\rm{dyn,eff}}^2}{(4\pi)^2} \left(\ln\frac{m_{\rm{dyn,eff}}^2}{4\pi\mu^2} -\psi(2) \right)+\frac{\left(\xi-1/6 \right)R}{(4\pi)^2}\left(\ln\frac{m_{\rm{dyn,eff}}^2}{4\pi\mu^2}-\psi(1)\right)\nonumber\\&& +\frac{ f_1}{(4\pi)^2m_{\rm{dyn,eff}}^2}+{\cal O}(\e).  
\label{v11}
\end{eqnarray}
where $d=4-\e$,  $\psi$ is the digamma function, and we have abbreviated
$$f_1 = \frac12\left(\frac16 -\xi\right)^2 R^2 -\frac13 a^{\lambda}\,_{\lambda} = \frac12\left(\frac16 -\xi\right)^2 {R^2} +\frac{1}{180}R_{\mu\nu\rho\lambda} R^{\mu\nu\rho\lambda} -\frac{1}{180} R_{\mu\nu} R^{\mu\nu}$$
For our purpose, we next abbreviate  \ref{v11} as,
\begin{equation}
G(x,x)= \left[m_{\rm{dyn,eff}}^2 + \left(\xi-\frac16  \right)R\right] f_{d}+ f_{\rm{fin}}
\label{v12}
\end{equation}  
where $f_d=- \mu^{-\e}/8\pi^2\e$, and
\begin{eqnarray}
&&f_{\rm{fin}} =  \frac{m_{\rm{dyn,eff}}^2}{(4\pi)^2} \left(\ln\frac{m_{\rm{dyn,eff}}^2}{4\pi\mu^2} -\psi(2) \right)+\frac{\left(\xi-1/6\right)R}{(4\pi)^2}\left(\ln\frac{m_{\rm{dyn,eff}}^2}{4\pi\mu^2} -\psi(1)\right)+\frac{ f_1}{(4\pi)^2m_{\rm{dyn,eff}}^2}
\label{v25}
\end{eqnarray}
Substituting \ref{v12} into \ref{v10}, we have
\begin{equation}
m_{\rm{dyn,eff}}^2+\xi R=m_0^2+\delta m_2^2+(\xi+\delta \xi_2)R+\frac{(\lambda+\delta \lambda_2) v^2}{2} +\frac{(\lambda+\delta \lambda_3)}{2}  \left[ \left(m_{\rm{dyn,eff}}^2 + \left(\xi-\frac16 \right)R \right)f_{d} + f_{\rm{fin}} \right].
\label{v13}
\end{equation}
From the above expression, we infer the most natural expression for the effective, dynamical mass squared
\begin{equation}
m_{\rm{dyn,eff}}^2=m_0^2+\frac{\lambda v^2}{2} +\frac{ \lambda f_{\rm{fin}}}{2}.
\label{v14}
\end{equation}
\ref{v25} shows that the above is a transcendental equation for $m_{\rm{dyn,eff}}^2$ and in general it can be found via numerical analysis. However, in the special case when $m_{\rm{dyn,eff}}^2/\mu^2$ is small, we find an approximate but otherwise non-perturbative expression,
\be
m^2_{\rm dyn, eff}\simeq  \frac{m_0^2+\lambda v^2/2 - \lambda (\xi-1/6)R\psi(1)/32\pi^2+\left[(m_0^2+\lambda v^2/2-\lambda (\xi-1/6)R\psi(1)/32\pi^2)^2 + \lambda f_1/8\pi^2\right]^{1/2}}{2},
\label{v15p}
\ee
where $f_1$ is defined below \ref{v11}.  In particular, for $v=0=m_0^2$, we have 
\be
m^2_{\rm dyn, eff}\simeq \frac{\sqrt{\lambda f_1}}{4\sqrt{2}\pi} \left(1 +{\cal O}(\sqrt{\lambda})\right),
\label{mda}
\ee
showing leading order field rest mass generation out of the spacetime curvature. For the de Sitter spacetime for example, we have $R=4\Lambda, \ R^{\mu\nu}R_{\mu\nu}=4\Lambda^2, \ R^{\mu\nu\rho\sigma}R_{\mu\nu\rho\sigma}=8\Lambda^2/3 $. Then  for minimal coupling ($\xi=0$), we have
$$\frac{m^2_{\rm dyn, eff}}{H^2}\simeq 0.9832 \times \frac{\sqrt{\lambda} }{4\pi}$$
where $H^2=\Lambda/3$. The above is approximately half the result found via using the exact large scale de Sitter propagator, e.g.~\cite{Bhattacharya:2025jtp} and references therein. This mismatch is expected, for as we have stated earlier, the present formalism  is UV effective and cannot capture the deep infrared effect at super-Hubble scales, unlike  the exact propagator.

We also note that in a maximally symmetric spacetime, one usually expects that the propagator in the coincident limit will be independent of the spacetime coordinates. This gives us the opportunity to perform resummation of the  coincident bubble self energy into a rest mass term, as has been done above. The de Sitter, being maximally symmetric, yields a self energy which is independent of the spacetime coordinates. However, we note  that the above prescription, being local, is valid for non-maximally symmetric spacetimes as well, simply because the curvature terms appearing in the Schwinger-DeWitt expansion are constant coefficients evaluated  at the centre of the normal coordinate system (cf., discussion made in \ref{S2'}). Thus should we wish to apply it for say, the Schwarzschild spacetime, we must substitute $R=0=R_{\mu\nu}$ and $R_{\mu\nu\lambda \rho}R^{\mu\nu\lambda \rho}=48 G^2M^2/r_0^6$ above, where $r_0$ is the radial point around which the normal coordinate has been erected in a sufficiently small neighbourhood, over which the variation of the curvature can be ignored.  In other words, \ref{mda} is always independent  of spacetime coordinates. The  difference between the maximally symmetric and other spacetimes would certainly be the fact that for the first, the dynamical mass will be a fixed quantity, whereas for the latter, its value would depend upon the point where the normal coordinate is erected. For example for the Schwarzschild, the dynamical mass would vanish for $r_0 \gg 2GM$.

\noindent
Finally, we also note the high field limit of the effective dynamical mass
\begin{eqnarray}
&&m_{\rm{dyn,eff}}^{2} \simeq  \frac{\lambda v^2}{2} +\frac{\lambda^2 v^2}{4(4\pi)^2} \ln\frac{v^2}{8\pi\mu^2} 
\label{v28'}
\end{eqnarray}

\noindent
The counterterms of \ref{v13} must satisfy the relationship
\begin{equation}
\delta m_2^2+\delta \xi_2 R+\frac{\delta \lambda_2 v^2}{2} +\frac{(\lambda+\delta \lambda_3) }{2} \left(m_{\rm{dyn,eff}}^2 + \left(\xi-\frac16 \right)R \right)f_{d}+\frac{\delta \lambda_2 f_{\rm{fin}}}{2} =0.
\label{v15}
\end{equation}
By collecting the coefficients of $v^0$, $v^2$, $R$ and $f_{\rm fin}$, and setting them to zero, the above equation is consistently solved by the following non-perturbative counterterms
\begin{eqnarray}
&&\delta \lambda_2 =\delta \lambda_3, \qquad
\lambda +\delta \lambda_2 = \lambda  \left(1+\frac{\lambda f_d}{2}\right)^{-1} \nonumber\\&&
\delta m_2^2  = -\frac{m_0^2 \lambda f_d}{2} \left(1+\frac{\lambda f_d}{2}\right)^{-1}, \qquad 
\delta \xi_2  =- \frac{\left(\xi-1/6 \right) \lambda f_{d}}{2}  \left(1+\frac{\lambda f_d}{2}\right)^{-1}.
\label{v16}
\end{eqnarray}

Likewise, the field equation \ref{v9} is self-renormalised and it should take the form,
\begin{eqnarray}
&&\left[-\Box +m_{\rm{dyn,eff}}^2+\xi R-\frac{\lambda v^2}{3} \right] v=0.
\end{eqnarray}
Using \ref{v16}, similar procedure as earlier then yields the consistent renormalisation conditions
\be
\delta m_1^2= \delta m_2^2, \qquad \delta \xi_1=\delta \xi_2, \qquad \delta \lambda_1 =3\delta \lambda_2
\label{CT}
\ee
The analysis so far determines the non-perturbative propagator in the Hartree approximation completely. We next substitute \ref{v8} into the 2PI effective action \ref{v4}, to find 
\begin{eqnarray}
&&\Gamma_{\rm 2PI}[v,iG]=-\! \int \! \! d^d x \sqrt{-g} \left[- \frac12 v \Box v+ \frac{1}{2} (m_0^2+\delta m_1^2+\left(\xi_1+\delta \xi_1 )R\right)v^2+\frac{(\lambda+\delta \lambda_1) v^4}{4!}  \right] \nonumber\\&&
-\frac{1}{2} \! \int \! \! d^d x \sqrt{-g} \! \int \! \! dm_{\rm{dyn,eff}}^2 \ iG(x,x)+ \frac{(\lambda+\delta \lambda_2)}{2^3}  \int \! \! d^d x \sqrt{-g}  (iG (x,x))^2.
\label{v17}
\end{eqnarray}
Using \ref{v12}, we next compute
\begin{eqnarray}
&&\int dm_{\rm{dyn,eff}}^2 iG(x,x)= \left(\frac{m_{\rm{dyn,eff}}^2}{2} + \left(\xi-\frac16  \right)R\right) m_{\rm{dyn,eff}}^2 f_{d} + \int dm_{\rm{dyn,eff}}^2 f_{\rm{fin}} \nonumber\\&& = \frac{m_0^4 f_d}{2} + \left(\xi -\frac16\right)R m_0^2 f_{d}+ \left(m_0^2 + \left(\xi-\frac16 \right)R \right)\frac{\lambda f_d v^2}{2} + \frac{\lambda^2 f_d v^4}{8}  + \frac{\lambda^2 f_d v^2}{4}  f_{\rm{fin}} \nonumber\\&& + \left( m_0^2  + \left(\xi -\frac16 \right)R \right)\frac{\lambda f_d  f_{\rm{fin}}}{2} + \frac{\lambda^2 f_d}{8} f_{\rm{fin}}^{2}+ \int dm_{\rm{dyn,eff}}^2 f_{\rm{fin}}
\label{v18}
\end{eqnarray}
and
\begin{eqnarray}
&&(iG(x,x))^2= \left[\left(m_{\rm{dyn,eff}}^2  + \left(\xi-\frac16 \right)R \right) f_{d} + f_{\rm{fin}}\right]^2 = m_0^4 f_{d}^{2} + \left(\frac16 -\xi \right)^2R^2 f_{d}^2 +\frac{\lambda^2 f_{d}^{2} v^4}{4} \nonumber\\&& + \left( m_0^2  +  \left(\xi-\frac16  \right)R  \right)\lambda f_d^2 v^2 + \left(\frac{\lambda f_{d}}{2}+1 \right)\lambda f_dv^2 f_{\rm{fin}} + \left(\frac{\lambda^2 f_{d}^{2}}{4}+\lambda f_d +1\right)f_{\rm{fin}}^{2} \nonumber\\&& +2\left(m_0^2 + \left(\xi-\frac16\right)R \right)\left( \frac{\lambda f_d}{2} +1\right)f_{d}f_{\rm{fin}}
\label{v19}
\end{eqnarray}
We  substitute \ref{v18}, \ref{v19} into \ref{v17}. Collecting the coefficients of  $v^2$, $v^4$, 
 $f_{\rm{fin}}$, $f_{\rm{fin}}^{2}$ and $v^2 f_{\rm{fin}}$, and using  
 \ref{v16} and \ref{CT}, we have
\begin{eqnarray}
&&v^2 : -\frac{\delta m_1^2}{2}-\frac{\delta \xi_1 R}{2}-\frac{\lambda f_d}{4} \left(m_0^2  + \left(\xi-\frac16\right)R \right)+\frac{\lambda (\lambda +\delta \lambda_2)f_d^2}{8} \left(m_0^2  +  \left(\xi-\frac16  \right)R  \right)=0
\nonumber\\
&&v^4: -\frac{\delta \lambda_2}{8}-\frac{\lambda^2 f_d}{16}+\frac{(\lambda +\delta \lambda_2)\lambda^2 f_{d}^{2}}{32}= 0 \nonumber\\
&& f_{\rm{fin}}: -\frac{\lambda f_d}{4} \left(m_0^2 + \left(\xi-\frac16 \right)R\right)+\frac{(\lambda +\delta \lambda_2)f_d}{4} \left(m_0^2  + \left(\xi-\frac16 \right)R \right)\left(\frac{\lambda f_d}{2}+1 \right)=0  \nonumber\\
&&f_{\rm{fin}}^{2}: -\frac{\lambda^2 f_d}{16}+\frac{(\lambda +\delta \lambda_2)}{8}\left(\frac{\lambda^2 f_{d}^{2}}{4}+\lambda f_d +1\right) =\frac{\lambda}{8}\nonumber\\
&&v^2 f_{\rm{fin}} : -\frac{\lambda^2 f_d}{8}+\frac{\lambda f_d(\lambda +\delta \lambda_2)}{8}\left(\frac{\lambda f_{d}}{2}+1 \right)=0.
\label{v20}
\end{eqnarray}
Note that after explicitly using the expression of the Green function, \ref{v17} becomes a function of the background field $v$ and curvature only.  Thus after such use, it should be  regarded as the 1PI effective action. Using \ref{v20}, the effective action becomes
\begin{eqnarray}
&&\Gamma_{\rm 1PI}[v]=-\! \int \! \! d^4 x \sqrt{-g} \Bigl[-\frac12 v \Box v+ \frac{1}{2} (m_0^2+\xi R)v^2+\frac{\lambda v^4}{4!}-\frac{\lambda f^{2}_{\rm{fin}}}{8}  +\frac12 \int dm_{\rm{dyn,eff}}^2 f_{\rm{fin}} \Bigr] \nonumber\\&&
+ \! \int \! \! d^d x \sqrt{-g} \left[-\frac{m_0^4 f_d}{4} - \left(\xi-\frac16 \right) \frac{R m_0^2 f_{d}}{2}+\frac{(\lambda +\delta \lambda_2)}{8}\left(m_0^4 f_{d}^{2} + \left(\xi-\frac16 \right)^2R^2 f_{d}^2\right)  \right].
\label{v21}
\end{eqnarray}
The first line of the above equation is free of any ultraviolet divergences. The second line can be absorbed in the gravitational counterterms defined in \ref{v6},
\begin{eqnarray}
&&\delta \Lambda = -\frac{m_0^4 f_d}{8}+\frac{(\lambda +\delta \lambda_2)}{16}m_0^4 f_{d}^{2},\qquad
\delta \kappa = -\left(\xi -\frac16 \right)\frac{m_0^2 f_{d}}{2}, \qquad
\delta \alpha = -\frac{(\lambda +\delta \lambda_2)}{8}\left(\xi-\frac16 \right)^2f_{d}^2 \nonumber\\
&& \delta \beta =0 =\delta \gamma. 
\label{v22}
\end{eqnarray}
This yields finally the renormalised effective action
\begin{eqnarray}
\Gamma[v]_{\rm 1PI, Ren.}=-\! \int \! \! d^4 x \sqrt{-g} \left[ -\frac12 v \Box v+\frac{1}{2} (m_0^2+\xi R)v^2+\frac{\lambda v^4}{4!}-\frac{\lambda f^{2}_{\rm{fin}}}{8}  +\frac12 \int dm_{\rm{dyn,eff}}^2 f_{\rm{fin}} \right]
\label{v23}
\end{eqnarray}

\noindent
Before we proceed further to investigate the effective potential, let us emphasise on the non-perturbative characteristics of the current formalism and \ref{v23}, which we also mentioned towards the end of the preceding Section.  First of all, the 2PI formalism is non-perturbative as it resums the self energy contributions originating from the {\it exact} propagator in the Hartree approximation. \ref{v14} shows the self energy contribution to the effective dynamical mass, as $\lambda f_{\rm fin}/2$ is the finite part of the one loop bubble self energy. Also, \ref{v11} and \ref{v25} show that $m^2_{\rm dyn, eff}$ is a non-trivial function of spacetime curvature, and it is clear  that $iG(x,x)$, \ref{v11}, resums the linear and quadratic order curvature terms. This can be thought of analogous to standard resummation of self energy in flat spacetimes,
$$\frac{1}{k^2+m_0^2} \to \frac{1}{k^2+m_0^2+i\Sigma}$$
where $i\Sigma$ is the self energy. In our case the self energy is a function of spacetime curvature as well, leading to resummation of the same and hence a {\it both way} non-perturbative result.

\subsection{The effective potential, and  
spontaneous symmetry breaking}\label{S4}
\begin{figure}[H]
\begin{center}
\includegraphics[scale=0.45]{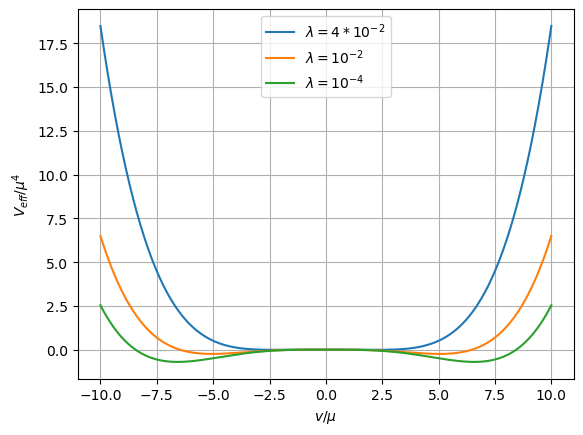}
\caption{  \it \small Plot of the effective potential \ref{v28'}, which is the flat spacetime and massless limit of the non-perturbative result of \ref{v28}.  We have made the potential dimensionless by scaling with respect to the renormalisation scale. This is basically the standard Coleman-Weinberg potential and shows spontaneous symmetry breaking feature. }
\label{symb1}
\end{center}
\end{figure}

\noindent
From the renormalised result \ref{v23}, we read off the effective potential,
\begin{eqnarray}
V_{\rm{eff}}(v)=\frac{1}{2} (m_0^2+\xi R)v^2+\frac{\lambda v^4}{4!}-\frac{\lambda f^{2}_{\rm{fin}}}{8}  +\frac12 \int dm_{\rm{dyn,eff}}^2 f_{\rm{fin}} 
\label{v24}
\end{eqnarray}
Using \ref{v25} we compute
\begin{eqnarray}
&&\int dm_{\rm{dyn,eff}}^2 f_{\rm{fin}}=\frac{m_{\rm{dyn,eff}}^4}{2(4\pi)^2} \left( \ln\frac{m_{\rm{dyn,eff}}^2}{4\pi\mu^2}- \frac12 -\psi(2) \right)+\frac{\left(\xi-1/6\right)R m_{\rm{dyn,eff}}^2 }{(4\pi)^2}\left(\ln\frac{m_{\rm{dyn,eff}}^2}{4\pi\mu^2} -\psi(2) \right) + f_1 \ln{\frac{m_{\rm{dyn,eff}}^2}{4\pi\mu^2}},\nonumber\\&&
\label{v27}
\end{eqnarray}
and the effective potential \ref{v24} explicitly becomes
\begin{eqnarray}
&&V_{\rm{eff}}(v)=\frac{1}{2} (m_0^2+\xi R)v^2+\frac{\lambda v^4}{4!}\nonumber\\&&-\frac{\lambda}{8} \left[\frac{m_{\rm{dyn,eff}}^2}{(4\pi)^2} \left(\ln\frac{m_{\rm{dyn,eff}}^2}{4\pi\mu^2} -\psi(2) \right)+\frac{\left(\xi-\frac16 \right)R}{(4\pi)^2}\left( \ln\frac{m_{\rm{dyn,eff}}^2}{4\pi\mu^2}-\psi(1)\right)+\frac{f_1}{(4\pi)^2m_{\rm{dyn,eff}}^2}\right]^2 \nonumber\\&& +\frac12 \left[\frac{m_{\rm{dyn,eff}}^4}{2(4\pi)^2} \left( \ln\frac{m_{\rm{dyn,eff}}^2}{4\pi\mu^2}- \frac12 -\psi(2) \right)+\frac{\left(\xi-\frac16\right)R m_{\rm{dyn,eff}}^2 }{(4\pi)^2}\left(\ln\frac{m_{\rm{dyn,eff}}^2}{4\pi\mu^2} -\psi(2) \right)+ \frac{f_1}{(4\pi)^2} \ln{\frac{m_{\rm{dyn,eff}}^2}{4\pi\mu^2}}\right]
\label{v28}
\end{eqnarray}
As a check of consistency, let us first take the massless, and zero curvature limit. In this case we simply have $m^2_{\rm dyn, eff}=  \lambda v^2/2$, to give
\begin{eqnarray}
&&V_{\rm{eff}}(v)\vert_{\rm flat, massless} = \frac{\lambda v^4}{4!} + \frac{\lambda^2 v^4}{2^8 \pi^2} \left(\ln \frac{v^2}{8\pi \mu^2} -\frac12 -\psi(2)\right)+{\cal O}(\lambda^3)
\label{v28'}
\end{eqnarray}
The above is the Coleman-Weinberg potential showing spontaneous symmetry breaking feature as shown in \ref{symb1}.

\begin{figure}[H]
\begin{center}
\end{center}
\centering
\begin{subfigure}{.4\textwidth}
  \centering
  \includegraphics[scale=0.25]{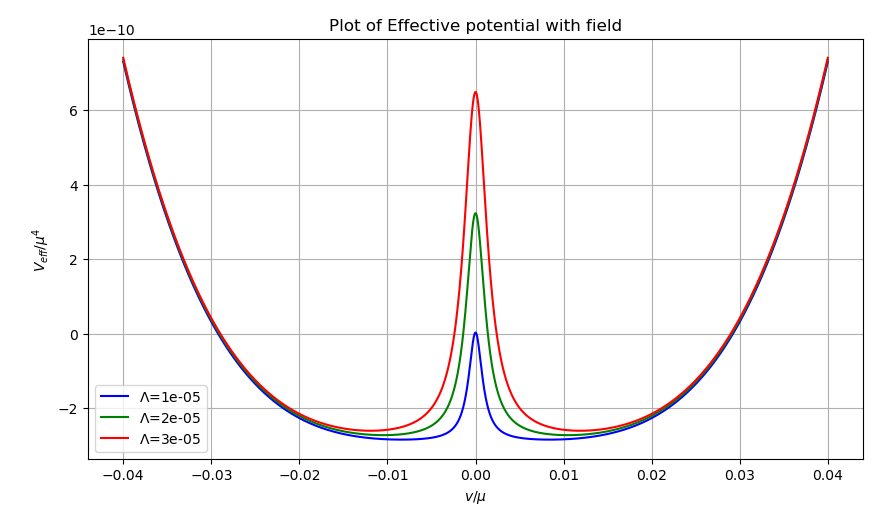}
  \caption{\small \it Smaller values of $\Lambda$.}
  \label{fig:sub7}
\end{subfigure}%
\begin{subfigure}{.7\textwidth}
  \centering
  \includegraphics[scale=0.25]{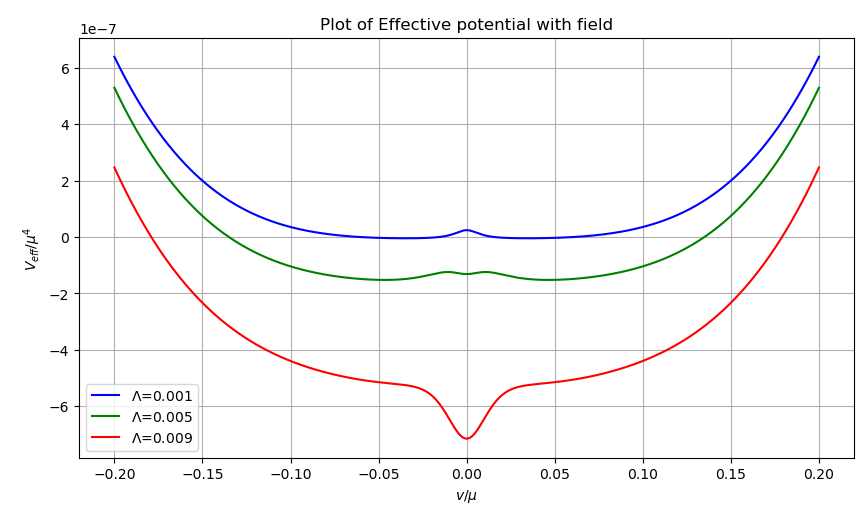}
  \caption{\small \it Relatively larger values of $\Lambda$.}
  \label{fig:sub8}
\end{subfigure}
\caption{\it \small Variation of  the non-perturbative effective potential, \ref{v28}, with respect to the background field. We have taken $m_0=10^{-5} \rm{GeV}$, and $\lambda =0.01$. $\Lambda$ is taken in ${\rm GeV}^2$. The left plot shows SSB pattern, whereas the right shows washing away of it via building a dip at the centre, with increasing $\Lambda$ values. See main text for discussion.}
\label{symb5}
\end{figure}

\noindent
Let us now come to \ref{v28}. We may explicitly evaluate it for any spacetime we wish. We will focus on the de Sitter spacetime in this work ($R=4\Lambda, \ R^{\mu\nu}R_{\mu\nu}=4\Lambda^2, \ R^{\mu\nu\rho\sigma}R_{\mu\nu\rho\sigma}=8\Lambda^2/3 $).   We have  plotted its variation in \ref{symb5}, for $m_0=10^{-5} \rm{GeV}$ and various values of  the cosmological constant, $\Lambda$. We have taken the scalar to be minimally coupled ($\xi=0$), so that the curvature contribution to the effective potential comes solely from the loop effects. Note that we have scaled $V_{\rm eff}$ by $\mu^4$. Thus the qualitative characteristics of this dimensionless effective potential will not depend upon the value of $\mu$. Also, we have taken a range for the values of  $\Lambda$, from $10^{-5}\rm{GeV^2}$ to  $0.01\rm{GeV^2}$. The quartic coupling $\lambda$ is taken to be $0.01$.

The first set of \ref{symb5}, which corresponds to comparatively lesser values of $\Lambda$, shows clear spontaneous  symmetry breaking (SSB) pattern. The second set of \ref{symb5}, which corresponds to relatively larger values of $\Lambda$, shows that the SSB is washed out with increasing $\Lambda$, and instead a minimum is developed at the centre.  The SSB found here for a massive field is curvature driven and was also present in the perturbative computations of~\cite{Nath:2024doz}. Note that {\it no} SSB is possible in flat spacetime, \ref{v28'}, if we add with it a mass term ($m_0^2 v^2/2$) with  $m_0^2\ {\ensuremath >} \ 0$. As we have stated earlier, the SSB found in~\cite{Nath:2024doz} was in quadratic order curvature computation, and the same was not present in earlier computations done in linear order  in the spacetime curvature. Even though \ref{w11} is written up to the quadratic order in the curvature, the 2PI technique sums those terms via the effective dynamical mass, \ref{v14}. Evidently, the curvature terms  appearing in the denominators of \ref{w11} can basically be expanded into  an infinite binomial  series.  Also, the  non-perturbative  effects seems to be reflected in obtaining  the SSB pattern with $\xi=0$, which was not possible in~\cite{Nath:2024doz}, at least for reasonable parameter values. Moreover,  the patterns in the second of~\ref{symb5} was also absent there. 
  Putting these all in together now, we conclude that curvature driven SSB is indeed possible for fields with positive rest mass squared, which has {\it no} analogue in flat spacetimes. We believe this result to be interesting in its own right.\\
  
 \noindent
 We now wish to extend the above result at high temperature. In particular, we wish to demonstrate the restoration of the symmetry. 

\section{Effective potential at finite temperature}\label{S5}
In the thermal case, \ref{v8}, \ref{v9} remains the same formally, with the propagator replaced by the thermal one,
\begin{equation}
 \left[- \Box_x+m_0^2+\delta m_{2}^{2}+(\xi+\delta \xi_2) R+\frac{(\lambda+\delta \lambda_2) v^2}{2} +\frac{(\lambda+\delta \lambda_3)}{2}  iG_{\beta}(x,x) \right] iG_{\beta}(x,x')
=- \frac{i\delta^d (x-x')}{\sqrt{-g}},
\label{t1}
\end{equation}
and
\begin{equation}
\left[- \Box+m_0^2+\delta m_{1}^{2}+(\xi+\delta \xi_1) R+\frac{(\lambda+\delta \lambda_1) v^2}{6} +\frac{ (\lambda+\delta \lambda_2)}{2}i G_{\beta}(x,x) \right] v(x)=0.
\label{v9t}
\end{equation}
\ref{t1} then suggests analogous equation as of \ref{v11'}, but with a temperature dependent effective dynamical mass,
\begin{eqnarray}
&& \left[-\Box_x+m_{\mathrm{\beta dyn,eff}}^2+\xi R\right] iG_{\beta}(x,x')=-i \delta^4 (x-x'),
\label{t2}
\end{eqnarray}
so that we have the relationship
\begin{equation}
m_{\beta\rm{dyn,eff}}^2+\xi R=m_0^2+\delta m_2^2+(\xi+\delta \xi_2) R+\frac{(\lambda+\delta \lambda_2) v^2}{2} +\frac{(\lambda+\delta \lambda_3) }{2} iG_{\beta}(x,x).
\label{t3}
\end{equation}
Let us  evaluate now $iG_{\beta}(x,x)$. In a thermal field theory, one may break the full  propagator into zero and finite temperature parts~\cite{Ashok},
\begin{eqnarray}
    &&\int\frac{d^dk}{(2\pi)^d}\frac{1}{(k^2 +m_{\beta\rm dyn, eff}^{2})} \rightarrow \int\frac{d^dk}{(2\pi)^d}\frac{1}{(k^2 +m_{\beta\rm dyn, eff}^{2})}+\int \frac{d^{3}\vec k}{(2\pi)^{3}} \frac{1}{\left(\vec{k}^2+m_{\beta\rm dyn, eff}^2\right)^{\frac12}}\frac{1}{e^{\beta\sqrt{\vec{k}^2+m_{\beta\rm dyn, eff}^2}} -1}\nonumber\\&&=\int\frac{d^dk}{(2\pi)^d}\frac{1}{(k^2 +m_{\beta\rm dyn, eff}^{2})}+S_{1}(\beta) \qquad ({\rm say}),
\label{v29}
\end{eqnarray}
The first part is the usual zero temperature propagator whereas the second, free of any ultraviolet divergence,  is the finite temperature part. Likewise, the other relevant integrals in the finite temperature version of \ref{v11} are
\begin{eqnarray}
&& \int \frac{d^dk}{(2\pi)^d}\left[ \frac{\left(1/6 -\xi \right)R}{(k^2 +m_{\beta\rm dyn, eff}^{2})^2} + \frac{2f_1}{(k^2 +m_{\beta\rm dyn, eff}^{2})^3}\right]\rightarrow \int \frac{d^dk}{(2\pi)^d}\left[ \frac{\left(1/6 -\xi \right)R}{(k^2 +m_{\beta\rm dyn, eff}^{2})^2} + \frac{2f_1}{(k^2 +m_{\beta\rm dyn, eff}^{2})^3}\right] \nonumber\\ &&-\left(\xi-\frac16 \right)RS_{2}(\beta)+2f_1S_{3}(\beta) \qquad ({\rm say})
\label{v30}
\end{eqnarray}
where
\begin{eqnarray}
&&S_2(\beta)= \int \frac{d^{3}\vec k}{(2\pi)^{3}} \frac{1}{\left(\vec{k}^2+m_{\beta\rm dyn, eff}^{2}\right)}\frac{1}{e^{\beta\sqrt{\vec{k}^2+m_{\beta\rm dyn, eff}^{2}}} -1}, \qquad S_3(\beta)=\int \frac{d^{3}\vec k}{(2\pi)^{3}} \frac{2}{\left(\vec{k}^2+m_{\beta\rm dyn, eff}^{2}\right)^{\frac32}}\frac{1}{e^{\beta\sqrt{\vec{k}^2+m_{\beta\rm dyn, eff}^{2}}} -1}\nonumber\\
\label{v30a1}
\end{eqnarray}
We thus have
\begin{equation}
G_\beta(x,x)= \left[m_{\beta\rm{dyn,eff}}^2 + \left(\xi-\frac16  \right)R\right] f_{d}+ f_{\beta\rm{fin}},
\label{t5}
\end{equation}  
with
\begin{equation}
f_{\beta\rm{fin}} = f_{\rm fin}+S_1(\beta)+\left(\frac16 -\xi \right)RS_{2}(\beta)+2f_1S_{3}(\beta),
\label{t6}
\end{equation}  
where $f_{\rm fin}$ is the zero temperature part given by \ref{v25}, however with $m^2_{\rm dyn, eff}$ replaced by $m_{\beta\rm dyn, eff}^{2}$. Substituting now \ref{t5} into \ref{t3}, we have
\begin{equation}
m_{\beta\rm{dyn,eff}}^2+\xi R=m_0^2+\delta m_2^2+(\xi+\delta \xi_2)R+\frac{(\lambda+\delta \lambda_2) v^2}{2} +\frac{(\lambda+\delta \lambda_3)}{2}  \left[ \left(m_{\beta\rm{dyn,eff}}^2 + \left(\xi-\frac16 \right)R \right)f_{d} + f_{\beta\rm{fin}} \right].
\label{t7}
\end{equation}
From the above expression, we get
\begin{equation}
m_{\beta\rm{dyn,eff}}^2=m_0^2+\frac{\lambda v^2}{2} +\frac{ \lambda f_{\beta\rm{fin}}}{2}.
\label{t8}
\end{equation}
The counterterms of \ref{t7} must satisfy the relationship
\begin{equation}
\delta m_2^2+\delta \xi_2 R+\frac{\delta \lambda_2 v^2}{2} +\frac{(\lambda+\delta \lambda_3) }{2} \left(m_{\beta\rm{dyn,eff}}^2 + \left(\xi-\frac16 \right)R \right)f_{d}+\frac{\delta \lambda_2 f_{\beta\rm{fin}}}{2} =0.
\label{v15}
\end{equation}
 The above equation is {\it exactly} formally the same as the zero temperature equation, \ref{v15}.
Accordingly, by collecting the coefficients of $v^0$, $v^2$, $R$ and $f_{\beta\rm fin}$, and setting them to zero we can immediately see that the counterterms are temperature independent and exactly same as that of the zero temperature case. \\

\noindent
Thus the effective action  for the thermal case formally looks the same as of \ref{v24},
\begin{eqnarray}
V_{\beta\rm{eff}}(v)=\frac{1}{2} (m_0^2+\xi R)v^2+\frac{\lambda v^4}{4!}-\frac{\lambda f^{2}_{\beta\rm{fin}}}{8}  +\frac12 \int dm_{\beta \rm{dyn,eff}}^2 f_{\beta\rm{fin}} 
\label{v23a}
\end{eqnarray}

\noindent
Also, the effective dynamical mass at zero temperature, \ref{v14}, in this case becomes
\be
m^2_{\beta {\rm dyn, eff}}= m_0^2 +\frac{\lambda v^2}{2} + \frac{\lambda f_{\beta{\rm fin}}}{2}= m_0^2 +\frac{\lambda v^2}{2} + \frac{\lambda f_{\rm fin}}{2}+\frac{\lambda}{2}\left(S_{1}(\beta)-\left(\xi-\frac16  \right)RS_{2}(\beta)+2f_1S_{3}(\beta)\right)
\label{v23b}
\ee

\noindent
We now compute
\begin{eqnarray}
&&f_{\beta\rm{fin}}^{2}=\left[\frac{m_{\beta\rm{dyn,eff}}^2}{(4\pi)^2} \left(\ln\frac{\beta m_{\rm{dyn,eff}}^2}{4\pi\mu^2} -\psi(2) \right)+\frac{\left(\xi-1/6 \right)R}{(4\pi)^2}\left( \ln\frac{m_{\beta \rm{dyn,eff}}^2}{4\pi\mu^2}-\psi(1)\right)+\frac{f_1}{(4\pi)^2m_{\beta \rm{dyn,eff}}^2}\right]^2 \nonumber\\&& +2\left[\frac{m_{\beta \rm{dyn,eff}}^2}{(4\pi)^2} \left(\ln\frac{m_{\beta \rm{dyn,eff}}^2}{4\pi\mu^2} -\psi(2) \right)+\frac{\left(\xi-1/6\right)R}{(4\pi)^2}\left(\ln\frac{m_{\beta \rm{dyn,eff}}^2}{4\pi\mu^2}-\psi(1)\right)+\frac{ f_1}{(4\pi)^2m_{\beta \rm{dyn,eff}}^2}\right] \times \nonumber\\&& \left[S_{1}(\beta)-\left(\xi-\frac16 \right) RS_{2}(\beta) +2f_1  S_{3}(\beta) \right] +\left[S_{1}(\beta)-\left(\xi-\frac16  \right)RS_{2}(\beta)+2f_1S_{3}(\beta)\right]^2,
\label{v31}
\end{eqnarray}
and,
\begin{eqnarray}
&&\int dm_{\beta\rm{dyn,eff}}^2 f_{\beta\rm{fin}}=\frac{m_{\beta\rm{dyn,eff}}^4}{2(4\pi)^2} \left( \ln\frac{m_{\beta\rm{dyn,eff}}^2}{4\pi\mu^2}- \frac12 -\psi(2) \right)+\frac{\left(\xi-1/6 \right)R \ m_{\beta\rm{dyn,eff}}^2 }{(4\pi)^2}\left(\ln\frac{m_{\beta\rm{dyn,eff}}^2}{4\pi\mu^2} -\psi(2) \right)\nonumber\\&& + f_1 \ln{\frac{m_{\beta\rm{dyn,eff}}^2}{4\pi\mu^2}}+\int dm_{\beta\rm{dyn,eff}}^2 \left(S_{1}(\beta)-\left(\xi-\frac16  \right)RS_{2}(\beta)+2f_1S_{3}(\beta)\right)
\label{v33}
\end{eqnarray}
The above equation along with \ref{v31}, \ref{v23b}, while substituted into \ref{v23a} give the desired renormalised effective potential. \\

\noindent
Let us evaluate the effective potential at the high temperature limit, $\beta \ll 1$. From \ref{v29}, \ref{v30a1}, we can make the high temperature expansions
\begin{eqnarray}
&&S_1(\beta)= \frac{1}{24\beta^2}-\frac{m_{\beta\rm{dyn,eff}}}{8\pi \beta}+\frac{m_{\beta\rm{dyn,eff}}^2}{16\pi^2}\left(\psi(1) +\ln\frac{2\pi}{m_{\beta\rm{dyn,eff}} \beta}\right) +\frac{m_{\beta\rm{dyn,eff}}^4 \beta^2\zeta(3)}{256\pi^4} - \frac{m_{\beta\rm{dyn,eff}}^6 \beta^4\zeta(5)}{6144\pi^6} + \frac{5m_{\beta\rm{dyn,eff}}^8 \beta^6\zeta(7) }{196608\pi^8}+\cdots \nonumber\\
&&S_2(\beta)=\frac{1}{16\pi\beta m_{\beta\rm{dyn,eff}}}-\frac{1}{16\pi^2}\left(\psi(1) +\ln\frac{2\pi}{m_{\beta\rm{dyn,eff}} \beta}\right) - \frac{m_{\beta\rm{dyn,eff}}^2 \beta^2 \zeta(3)}{128\pi^4} +\frac{m_{\beta\rm{dyn,eff}}^4 \beta^4 \zeta(5)}{2048\pi^6}-\frac{5 m_{\beta\rm{dyn,eff}}^6 \beta^6 \zeta(7)}{49152\pi^8} + \cdots\nonumber\\
&&S_3(\beta)= \frac{1}{32\pi\beta m_{\beta\rm{dyn,eff}}^3}+\frac{\beta^2 \zeta(3)}{128\pi^4}-\frac{m_{\beta\rm{dyn,eff}}^2 \beta^4 \zeta(5)}{1024 \pi^6}+\frac{15m_{\beta\rm{dyn,eff}}^4\beta^6 \zeta(7)}{49152\pi^8}+\cdots
\label{v32}
\end{eqnarray}
Using the above, we compute for the integral on the right hand side of \ref{v33},
\begin{eqnarray}
\int dm_{\beta\rm{dyn,eff}}^2 \left[S_{1}(\beta)-\left(\xi-\frac16  \right)RS_{2}(\beta)+2f_1S_{3}(\beta)\right]\simeq S_{a}(\beta)-\left(\xi-\frac16 \right)RS_{b}(\beta)+2f_1S_{c}(\beta),
\label{v34}
\end{eqnarray}
where we have abbreviated,
\begin{eqnarray}
&&S_{a}(\beta)= \left(\frac{m_{\beta\rm{dyn,eff}}^2 }{24\beta^2}-\frac{m_{\beta\rm{dyn,eff}}^3}{12\pi \beta} -\frac{ m_{\beta\rm{dyn,eff}}^4 \ln{m_{\beta\rm{dyn,eff}}^2 \beta^2}}{64\pi^2}+\frac{m_{\beta\rm{dyn,eff}}^4\left(2\psi(1) +\ln{4\pi}\right)}{64\pi^2}   +\frac{m_{\beta\rm{dyn,eff}}^6\beta^2\zeta(3)}{768\pi^4}\right.\nonumber\\&&\left. -\frac{m_{\beta\rm{dyn,eff}}^8\beta^4\zeta(5)}{24576\pi^6} +\frac{m_{\beta\rm{dyn,eff}}^{10}\beta^6\zeta(7)}{196608\pi^8} \right)\nonumber\\
&&S_b(\beta)=\left(\frac{m_{\beta\rm{dyn,eff}}}{8\pi \beta}-\frac{m_{\beta\rm{dyn,eff}}^2}{16\pi^2}\left(\psi(1) +\ln{2\pi}\right) + \frac{m_{\beta\rm{dyn,eff}}^2}{32\pi^2}\ln{m_{\beta\rm{dyn,eff}}^2 \beta^2}-\frac{m_{\beta\rm{dyn,eff}}^4 \beta^2 \zeta(3)}{256\pi^4} \right. \nonumber\\ && \left. + \frac{m_{\beta\rm{dyn,eff}}^6 \beta^4\zeta(5)}{6144\pi^6}   - \frac{5m_{\beta\rm{dyn,eff}}^8 \beta^6\zeta(7)}{196608\pi^8}\right) \nonumber\\
&&S_c(\beta)= -\left(\frac{1}{16\pi\beta m_{\beta\rm{dyn,eff}}}- \frac{m_{\beta\rm{dyn,eff}}^2 \beta^2\zeta(3)}{128\pi^4} +\frac{m_{\beta\rm{dyn,eff}}^4 \beta^4\zeta(5)}{2048\pi^6}-\frac{5 m_{\beta\rm{dyn,eff}}^6 \beta^6\zeta(7)}{49152\pi^8} \right)
\label{v35}
\end{eqnarray}
Putting things together,  we now explicitly have the effective potential at finite temperature
\begin{eqnarray}
&&V_{\rm{eff,\beta}}=V_{\rm{eff}}  -\frac{\lambda}{8}\left[2\left(\frac{m_{\beta\rm{dyn,eff}}^2}{(4\pi)^2} \left(\ln\frac{m_{\beta\rm{dyn,eff}}^2}{4\pi\mu^2} -\psi(2) \right)+\frac{\left(\xi-1/6 \right)R}{(4\pi)^2}\left( \ln\frac{m_{\beta\rm{dyn,eff}}^2}{4\pi\mu^2}-\psi(1)\right)  +\frac{ f_1}{(4\pi)^2m_{\beta\rm{dyn,eff}}^2}\right)\right. \nonumber\\&& \left.\times \left(S_{1}(\beta)+\left(\frac16 -\xi \right) RS_{2}(\beta)  +2f_1  S_{3}(\beta) \right) +\left(S_{1}(\beta)-\left(\xi-\frac16  \right)RS_{2}(\beta) +2f_1S_{3}(\beta)\right)^2\right] \nonumber\\&& +\frac12 \left(S_{a}(\beta)-\left(\xi-\frac16 \right)RS_{b}(\beta)+2f_1S_{c}(\beta)\right),
\label{v36}
\end{eqnarray}
where $V_{\rm{eff}}$ is the same as the zero temperature effective potential, with $m^2_{\rm dyn, eff}$ replaced by $m^2_{\beta{\rm dyn, eff}}$.\\

\noindent
As a check of consistency,  we note that at extreme high temperature \ref{v23b}, \ref{v32} gives,
\be
m^2_{\beta{\rm dyn, eff}} \sim \lambda T^2
\label{dm}
\ee
which is well known in thermal field theory found via the hard thermal loop approximation, e.g.~\cite{Rajesh, Arjun} and references therein.\\

\begin{figure}[H]
\begin{center}
\includegraphics[scale=0.25]{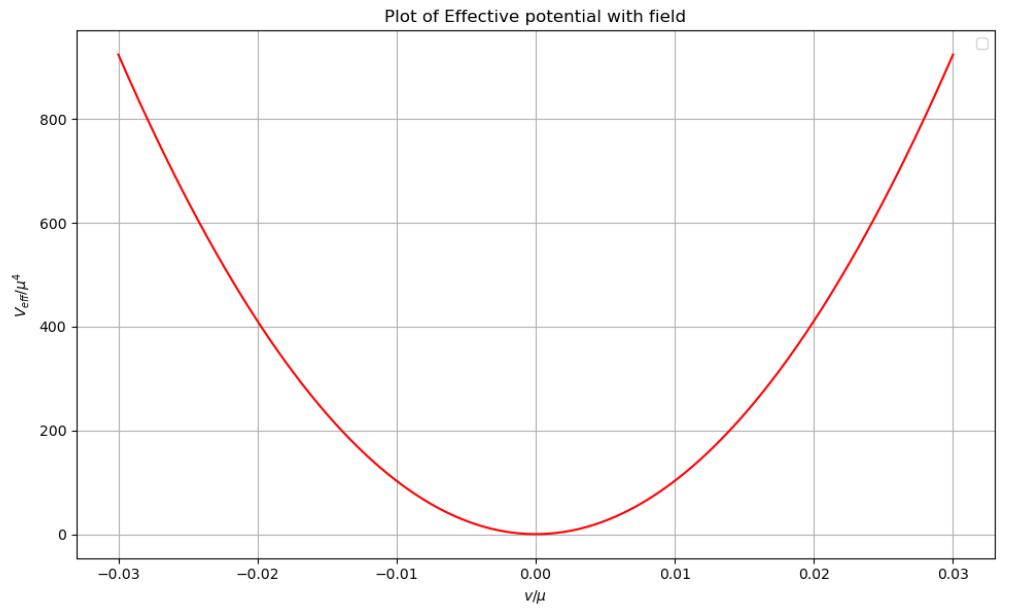}
\caption{  \it \small Plot of the high temperature effective potential, \ref{v36}, in the de Sitter background. There is no spontaneous symmetry breaking unlike the zero temperature case. We have taken $m_0= 10^{-5}{\rm GeV}$, $\lambda =0.01$, $\Lambda=10^{-5}{\rm GeV}^2$ and $\xi=0$.}
\label{symb4}
\end{center}
\end{figure}

\noindent
We have plotted the variation of \ref{v36} with respect to the background field for $\beta=10^{-5}{\rm{GeV}^{-1}}$ in \ref{symb4}, for the de Sitter spacetime. Unlike the zero temperature case, there is no spontaneous symmetry breaking here. This confirms symmetry restoration at high temperature. 

\noindent

Before we end this section, we wish to emphasise the importance of the dynamical mass in reaching the correct physical conclusion, by observing a little thing. Let us take the thermal part of the one loop self energy in the $\phi^4$ theory~\cite{Ashok},
\begin{eqnarray}
\Sigma = 12\lambda\int \frac{d^3\vec k}{(2\pi)^3} \frac{1}{\left(\vec{k}^2+m^2\right)^{\frac12}}\frac{1}{e^{\beta\sqrt{\vec{k}^2+m^2}} -1}.
\label{g1}
\end{eqnarray}
Let us now `turn on' the gravity  and for simplicity assume the spacetime to be weakly curved. Keeping only up to the ${\cal O}(R)$ term, we have for a light field~\cite{Nath:2024doz},
\begin{eqnarray}
\Sigma =\lambda T^2 -\frac{3\lambda m T}{\pi}+\left(\frac16 -\xi \right)\frac{3\lambda R T}{2\pi m}.
\label{g4}
\end{eqnarray}
Thus the curvature `correction' term blows up as $m\to 0$. However, if we take $m$ to be the effective dynamical mass, there is no problem in setting the field  rest mass to zero. \\

We recall that in a curved spacetime with isometries,  such as  in static spacetimes like the Schwarzschild or de Sitter, the frequency of a field gets redshifted as it climbs a gravitational hill up. The frequency is defined with respect to the Killing timelike parameter. In a thermal ensemble, the invariance of   the distribution function then implies a compensating blueshift of the temperature : $T \to T/\sqrt{|g_{00}|}$, known as the Tolman temperature. In the Schwinger-DeWitt expansion  however, the frequency of the field is {\it not} defined with respect to the timelike Killing vector field of the ambient spacetime, but with respect to the Riemann normal coordinate's time, $y^0$, of \ref{w4}. Second, in the Hartree approximation, we have evaluated things in the coincidence limit, $x\to x'$, of the propagator. By construction, $x-x'=0$ is  the origin of the normal coordinates, where $g_{\mu\nu}=\eta_{\mu\nu}$. Hence no notion of the Tolman temperature seems to be possible in our present case. However, this should not be surprising, for the gravitational redshift is essentially  a large scale phenomenon. It involves the observation of two persons large separated in the spacetime. The Schwinger-DeWitt expansion, as we already have emphasised earlier, is  short scale, involving observers located within a small neighbourhood.  Moreover, the Hartree approximation essentially involves evaluation of quantities at a single point, which happens to be the origin of the normal coordinates.   \\

\noindent
We would now like to extend the above results at zero and finite temperatures for an $O(N)$ symmetric scalar field theory.

\section{The case of an $O(N)$ scalar field theory}\label{ON0}
\subsection{The symmetric phase}\label{sphasel}

For an $O(N)$ linear sigma model scalar field theory in curved spacetime the action reads~\cite{Romatschke:2024hpb}
\begin{eqnarray}
S[\phi_i]=-\! \int  d^dx\sqrt{-g} \left[ \frac{1}{2} \phi^i (-\Box +m_0^2) \phi^i+\frac{1}{2}\xi R \phi^i\phi^i+\frac{\lambda}{4!N} (\phi^i\phi^i)^2 \right],
\label{v37}
\end{eqnarray}
where $i= 1, 2, \cdots N$. We make the decomposition of $\phi^i$ into classical background and quantum fluctuation,
$$\phi^i = v^i +\varphi^i,$$
so that the action becomes
\begin{eqnarray}
&& S[v^i+\varphi^i]
=
S[v^i]+\int d^d x\sqrt{-g} \left[  -\frac{1}{2}\varphi^i(-\Box+m_0^2+\xi R)\varphi^i -\frac{\lambda}{4!N}\left[2(\varphi^i\varphi^i)(v^jv^j) +4(\varphi^iv^i)^2\right] \right.\nonumber\\
&& \left.
-\frac{\lambda}{4!N}\left[(\varphi^i\varphi^i)^2+4(\varphi^i\varphi^i)(v^j\varphi^j)\right] \right].
\label{v38}
\end{eqnarray}
From the above action, we may write down the free inverse propagator and the interaction terms respectively as 
\begin{eqnarray}
i G_{0ij}^{-1}(x,x')=\frac{\delta^2 S[v^k]}{\delta v^i (x)\delta v^j(x') }
\label{v39}
\end{eqnarray}
\begin{eqnarray}
S_{\mathrm{int}}=-\frac{\lambda}{4!N}\! \int \! \! d^dx\sqrt{-g} \Bigl[ 2(\varphi^i\varphi^i)(v^jv^j)+4(\varphi^iv^i)^2\Bigr] 
-\frac{\lambda}{4!N}\! \int \! \! d^dx\sqrt{-g}\Bigl[ (\varphi^i\varphi^i)^2+4(\varphi^i\varphi^i)(v^j\varphi^j)\Bigr]
\label{v40}
\end{eqnarray}
\begin{figure}[H]
\begin{center}
\includegraphics[scale=0.50]{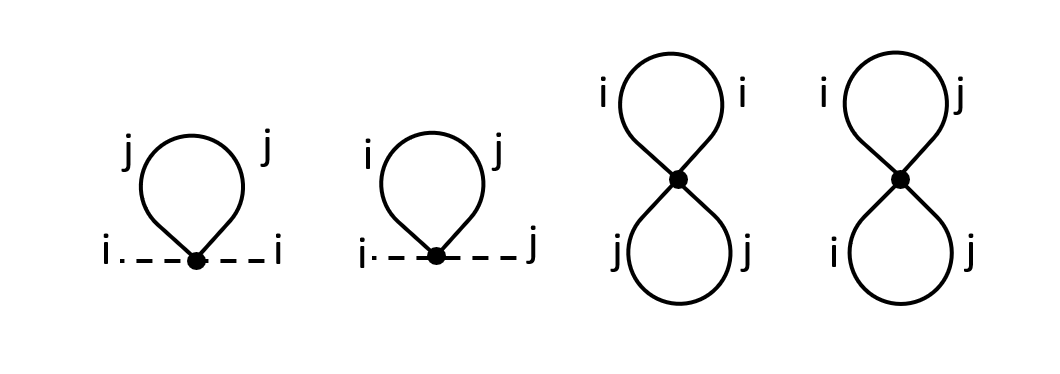}
\caption{  \it \small Lowest order Feynman diagrams pertaining to the symmetric phase of the $O(N)$ model, \ref{v40}. The solid line stands for the exact scalar propagators, whereas the dashed lines stand for the background field. Note that the leading contribution from the last term in \ref{v40} comes at ${\cal O}(\lambda^2)$, and hence is not included here for our present purpose. }
\label{symb101}
\end{center}
\end{figure}
The 2PI effective action corresponding to the above model is given by,
\begin{eqnarray}
    &&\Gamma_{\rm 2PI}[v^i,iG_{ij}]=S[v^i]-\frac{i}{2}\mathrm{Tr}\ln \det[iG_{ij}] +\frac{1}{2}\mathrm{Tr}\int d^dx d^d x' \sqrt{-g}   \sqrt{-g'}[iG_{0,ij}^{-1}[v^i](x,x')iG_{ji}(x',x)]+i\Gamma_2[v^i,iG_{ij}],
    \label{y1}
\end{eqnarray}
where $i\Gamma_2[v^i,iG_{ij}]$ represents the contributions from the 2PI vacuum diagrams. Under the Hartree approximation at 2 loop, the above effective action explicitly reads
\begin{eqnarray}
    &&\Gamma_{\rm 2PI}[v^i,iG_{ij}]=-\int d^dx \sqrt{-g}  \left[ \frac{1}{2}v^i(-\Box+m_0^2+\delta m_1^2 +(\xi +\delta \xi_1) R)v^i+\frac{\lambda+\delta\lambda_1}{4!N}(v^iv^i)^2 \right]-\frac{i}{2}\mathrm{Tr}\ln \det[iG_{ij}]\nonumber\\
&&-\frac{1}{2}\mathrm{Tr}\int d^dx  \sqrt{-g}   \left[ -\Box+m_0^2+\delta m_2^2+(\xi +\delta \xi_2) R+\frac{1}{6N}(\lambda+\delta\lambda_2^A)   v^kv^k+\frac{1}{3N}(\lambda+\delta\lambda_2^B) v^iv^j\right]iG_{ij}(x,x) \nonumber\\
&&-\frac{1}{4!N}\int \!d^d x \sqrt{-g} \left[ (\lambda+\delta\lambda_3^A)  iG_{ii}(x,x)  iG_{jj}(x,x)+2(\lambda+\delta\lambda_3^B)  (iG_{ij}(x,x))^2 \right],
\label{y2}
\end{eqnarray}
where we have used for the tree level inverse propagator \ref{v39},
\begin{eqnarray}
&&iG_{0,ij}^{-1}[v^i](x,x')=-\left[\left( -\Box+m_0^2+(\xi+\delta \xi_2) R+\frac{\lambda+\delta\lambda_2^A}{6N}(v^k)^2\right) \delta^{ij}+\frac{\lambda+\delta\lambda_2^B}{3N}v^iv^j\right]\delta^d (x,x')
\label{y1}
\end{eqnarray}
Compared to the $N=1$ case discussed earlier, there are two different kind of vertex counterterms of $A$ and $B$ type. This originates from the two different field orientations in the double bubble as shown in  \ref{symb101}. Putting $N=1$ and $\delta \lambda^A=\delta \lambda^B$, we reproduce the corresponding equations for the simple $\phi^4$ theory.\\

\noindent
The symmetric phase is characterised by $v^i=v$, for all $i=1,2, \cdots N$. Thus $v^i v^i = N v^2 $. Due to the $O(N)$ symmetry, $iG_{ij}(x,x')$ will simply be proportional to $\delta_{ij}$, $iG_{ij}(x,x')= iG(x,x')\delta_{ij}$ (say). We expect the scenario in this phase will be qualitatively similar to that of the simple $\phi^4$ theory.

Using now $\mathrm{Tr}(\delta_{ij}iG_{ij})=\mathrm{Tr}(iG_{ij})$, \ref{y2} simplifies to
\begin{eqnarray}
    &&\Gamma_{\rm 2PI}[v,iG]=-N\int d^d x \sqrt{-g} \left[ \frac{1}{2}v\left(-\Box+m_0^2+\delta m_1^2 + (\xi +\delta \xi_1 )R\right)v+\frac{(\lambda+\delta\lambda_1) v^4}{4!} \right] -\frac{i N}{2}\ln\det[iG(x,x)] \nonumber\\
&&-\frac{N}{2}\int d^d x \sqrt{-g} \left[(-\Box+m_0^2+\delta m_2^2 +(\xi +\delta \xi_2 )R+\frac16\left(\left(\lambda+\delta\lambda_2^A)+\frac{2}{N}(\lambda+\delta\lambda_2^B\right)\right)v^2 iG(x,x)\right]\nonumber\\
&& -\frac{N}{4!}\int \!d^dx \sqrt{-g} \left[\left(\lambda+\delta\lambda_3^A\right)+\frac{2}{N}\left(\lambda+\delta\lambda_3^B\right)\right](iG(x,x))^2.
\label{y3}
\end{eqnarray}
The equations of motion are given by,
\begin{equation}
\left[ -\Box+m_0^2+\delta m_1^2+ (\xi+\delta\xi_1)R+\frac{1}{6}(\lambda+\delta\lambda_1) v^2+\frac16\left(\left(\lambda+\delta\lambda_2^A\right)+\frac{2}{N}\left(\lambda+\delta\lambda_2^B\right)\right)iG(x,x)\right] v(x)=0,
\label{1d4}
\end{equation}
and
\begin{eqnarray}
&&-\left[ -\Box_x+m_0^2+\delta m_2^2+ (\xi+\delta\xi_2) R+\frac16\left( \left(\lambda+\delta\lambda_2^A\right)+\frac{2}{N}\left(\lambda+\delta\lambda_2^B\right) \right)v^2+\frac{1}{6}\left(\left(\lambda+\delta\lambda_3^A\right)+\frac{2}{N}\left(\lambda+\delta\lambda_3^B\right)\right)iG(x,x)\right] iG(x,x')\nonumber\\ &&=i\delta^d(x-x').
\label{y4g}
\end{eqnarray}
From the above equation, we have for consistent renormalisation 
\begin{equation}
m_{\rm{dyn,eff}}^2 + \xi R=m_0^2+ \delta m_2^2+(\xi+\delta \xi_2)R+\frac16 \left(\lambda + \delta \lambda_2^A+\frac{2}{N}\lambda+\frac{2}{N}\delta \lambda_2^B\right)v^2+\frac16 \left(\lambda + \delta \lambda_2^A+\frac{2}{N}\lambda+\frac{2}{N}\delta \lambda_2^B\right)iG(x,x) .
\label{y5}
\end{equation}
We now substitute \ref{v12} into the above equation. It turns out that for consistent renormalisation we must have
$$\delta \lambda^{A}_3=\delta \lambda^{A}_2, \qquad \delta \lambda^{B}_3=\delta \lambda^{B}_2, $$
so that,
\begin{equation}
m_{\rm{dyn,eff}}^2=m_0^2+\frac16 \left(1+\frac{2}{N}\right)\lambda v^2 +\frac16 \left(1+\frac{2}{N}\right)\lambda f_{\rm{fin}},
\label{v14s}
\end{equation}
where $f_{\rm fin}$ is given by \ref{v25}. Note that the effective dynamical mass decreases with increasing $N$.
Putting in $N=1$ above reproduces  \ref{v14}. We now read off  the following renormalisation condition,
\begin{eqnarray}
&&\delta m_2^2 + \delta\xi_2 R+ \frac16 \left(\delta \lambda_2^A +\frac{2}{N}\delta \lambda_2^B \right)v^2+\frac16 \left(\left(1+\frac{2}{N}\right)\lambda +\delta \lambda_2^A +\frac{2}{N}\delta \lambda_2^B  \right)\left[\left(m_0^2+\frac16 \left(1+\frac{2}{N}\right)\lambda v^2 +\frac16 \left(1+\frac{2}{N}\right)\lambda f_{\rm{fin}}\right)f_d \right. \nonumber\\&& \left. +\left(\xi -\frac16 \right)R f_{d} \right]+\frac16 \left(\delta\lambda_2^A +\frac{2}{N}\delta\lambda_2^B \right)f_{\rm{fin}} = 0
\label{v52}
\end{eqnarray}
We have from the above equation
\begin{eqnarray}
&& \delta \lambda_2^A +\frac{2}{N}\delta \lambda_2^B = -\frac16 \left(1+\frac{2}{N}\right)^2\lambda^2 f_d \left[1+\frac16 \left(1+\frac{2}{N}\right) \lambda f_d\right]^{-1} , \nonumber\\&&
 \left(1+\frac{2}{N}\right)\lambda +\delta \lambda_2^A +\frac{2}{N}\delta \lambda_2^B  = \left(1+\frac{2}{N}\right)\lambda \left[1+\frac16 \left(1+\frac{2}{N}\right) \lambda f_d\right]^{-1},  \nonumber\\&&
\delta m^2_2  = -\frac16  \left[\left(1+\frac{2}{N}\right)\lambda +\delta \lambda_2^A +\frac{2}{N}\delta \lambda_2^B \right] m_0^2 f_d ,\nonumber\\&& \qquad 
\delta \xi_2  = -\frac{\left(\xi-1/6 \right)  f_{d}}{6} \left[\left(1+\frac{2}{N}\right)\lambda +\delta \lambda_2^A +\frac{2}{N}\delta \lambda_2^B \right].
\label{v16'}
\end{eqnarray}
We could not find any further condition that tells us how individually $\delta \lambda_2^{A,B} $ look like. Hence instead we will keep the combination $\delta \lambda_2^A +2\delta \lambda_2^B/N$ as it is. This will not pose any problem in renormalising the effective action. Using now \ref{1d4}, we obtain further consistent renormalisation conditions
\begin{equation}
    \delta m_1^2=\delta m_2^2,~\delta\xi_1=\delta\xi_2,~\delta\lambda_1=\delta \lambda_2^A +\frac{2}{N}\delta \lambda_2^B
\end{equation}

\noindent
In order to derive  the effective action explicitly, we now compute
\begin{eqnarray}
&&\int dm_{\rm{dyn,eff}}^2 iG(x,x)
= \frac{f_d}{2}\left[m_0^4+ \frac{\lambda^2 \left(1+2/N\right)^2 v^4}{36}+\frac{\lambda^2 \left(1+2/N\right)^2 f_{\rm{fin}}^2}{36} \right. \nonumber\\&& \left.  +2m_0^2\left(\frac{\lambda \left(1+2/N\right) v^2}{6}+\frac16 \lambda \left(1+2/N\right)f_{\rm{fin}}\right)+\frac{\lambda^2 \left(1+2/N\right)^2 v^2 f_{\rm{fin}}}{18}\right]  \nonumber\\&&  + \left(\xi -\frac16  \right)Rf_d\left(m_0^2+\frac16 \left(1+\frac{2}{N}\right)\lambda v^2 +\frac16 \left(1+\frac{2}{N}\right)\lambda f_{\rm{fin}}\right)+ \int dm_{\rm{dyn,eff}}^2 f_{\rm{fin}},
\label{v18'}
\end{eqnarray}
and
\begin{eqnarray}
&&(iG(x,x))^2= \left[m_0^4+ \frac{\lambda^2 \left(1+2/N\right)^2 v^4}{36}+\frac{\lambda^2 \left(1+2/N\right)^2 f_{\rm{fin}}^2}{36} \right. \nonumber\\&& \left.  +2m_0^2\left(\frac{\lambda \left(1+2/N\right) v^2}{6}+\frac16 \lambda \left(1+\frac{2}{N}\right)f_{\rm{fin}}\right)+\frac{\lambda^2 \left(1+2/N\right)^2 v^2 f_{\rm{fin}}}{18}\right]f_d^2 \nonumber\\&& +2 \left[m_0^2+\frac16 \left(1+\frac{2}{N}\right)\lambda v^2+\frac16 \left(1+\frac{2}{N}\right)\lambda f_{\rm{fin}}\right]\times\left[\left(\xi -\frac16  \right)Rf_d+f_{\rm{fin}}\right]f_d+\left(\xi -\frac16 \right)^2R^2f^2_d\nonumber\\&& +2\left(\xi -\frac16  \right)Rf_d f_{\rm{fin}} +f_{\rm{fin}}^2. 
\label{v19'}
\end{eqnarray}
We next use \ref{y4g}  along with the above expressions into \ref{y3} and use the renormalisation conditions. This yields the effective action
\begin{eqnarray}
&&\Gamma_{\rm 1PI}[v]=-N\int \! \! d^4 x \sqrt{-g} \left[ \frac{1}{2} v\left(-\Box +  m_0^2+\xi R\right)v+\frac{\lambda v^4}{4!}-\frac{(1+ 2/N)\lambda f^{2}_{\rm{fin}}}{24}  +\frac12 \int dm_{\rm{dyn,eff}}^2 f_{\rm{fin}} \right] \nonumber\\&&
+ N \int \! \! d^d x \sqrt{-g}  \left[-\frac{m_0^4 f_d}{4} - \left(\xi -\frac16  \right)R \frac{m_0^2 f_{d}}{2} \right. \nonumber\\&& \left. +\frac{\left(\left(1+2/N\right)\lambda +\delta \lambda_2^A +2\lambda_2^B/N\right)}{24}\left(m_0^4 f_{d}^{2} + \left(\xi -\frac16  \right)^2R^2 f_{d}^2  +2m_0^2f_d \left(\xi  -\frac16 \right)R f_d\right)\right].
\label{v21}
\end{eqnarray}
The field independent divergences  appearing in the second and third lines can be cancelled via the  gravitational counterterms, \ref{v6}.  Hence we read off the renormalised effective potential from \ref{v21},
\begin{eqnarray}
V_{\rm{eff}}(v)=N\left[\frac12(m_0^2+\xi R)v^2+\frac{\lambda v^4}{4!}-\frac{\left(1+2/N\right)\lambda f^{2}_{\rm{fin}}}{24}  +\frac12 \int dm_{\rm{dyn,eff}}^2 f_{\rm{fin}} \right]
\label{v21'}
\end{eqnarray}
Putting $N=1$ above reproduces \ref{v24}. We note in particular that unless $N$ is large,  the $f_{\rm fin}^2$ term  has a different $N$-dependence from the rest. This is a non-trivial feature absent in the standard one loop effective action. In \ref{symb4}, we have investigated the behaviour of the effective potential with $N=2$ in the de Sitter background. We see that SSB is present, as expected in the symmetric phase. Also in \ref{symb5}, we have investigated the variation of $V_{\rm eff}(v)$ with respect to the $N$ values.   
\begin{figure}[H]
\begin{center}
\includegraphics[scale=0.30]{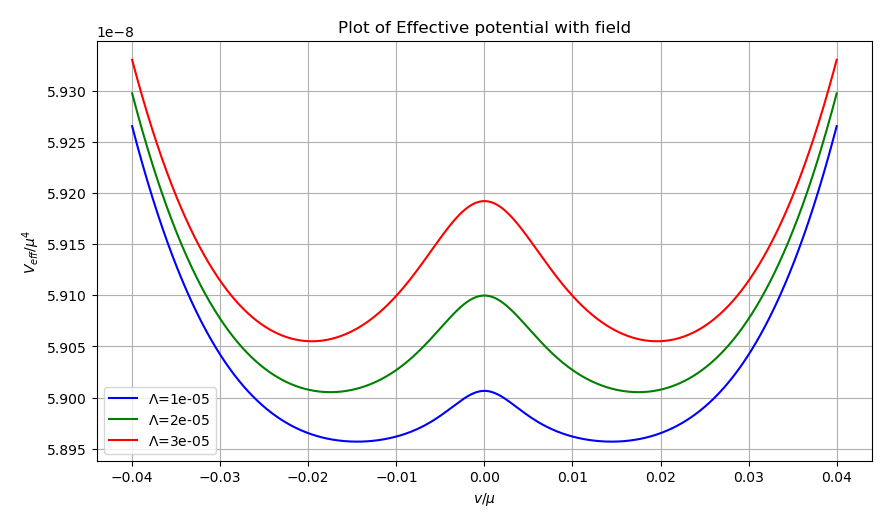}
\caption{  \it \small Variation of the effective potential, \ref{v21'}, with the background field for de Sitter spacetime with a couple of values of $\Lambda$ in ${\rm GeV}^2$,  $\lambda = 0.01$, $N=2$ and $m_0=10^{-5}${\rm GeV}. The SSB pattern is eminent. }
\label{symb4'}
\end{center}
\end{figure}
\begin{figure}[H]
\begin{center}
\includegraphics[scale=0.19]{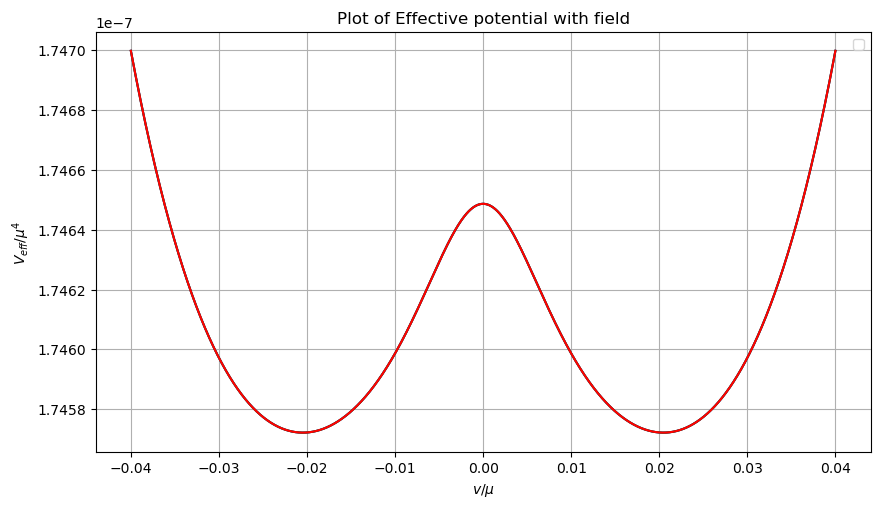}
\includegraphics[scale=0.19]{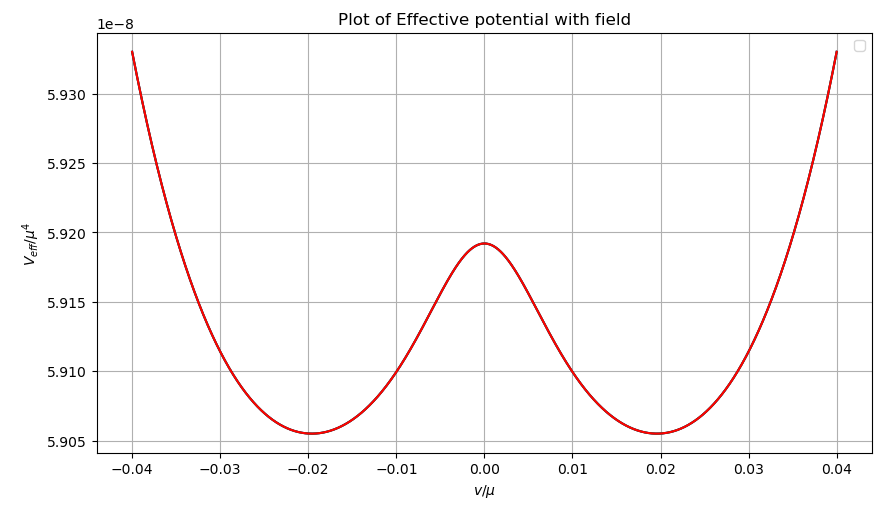}
\includegraphics[scale=0.19]{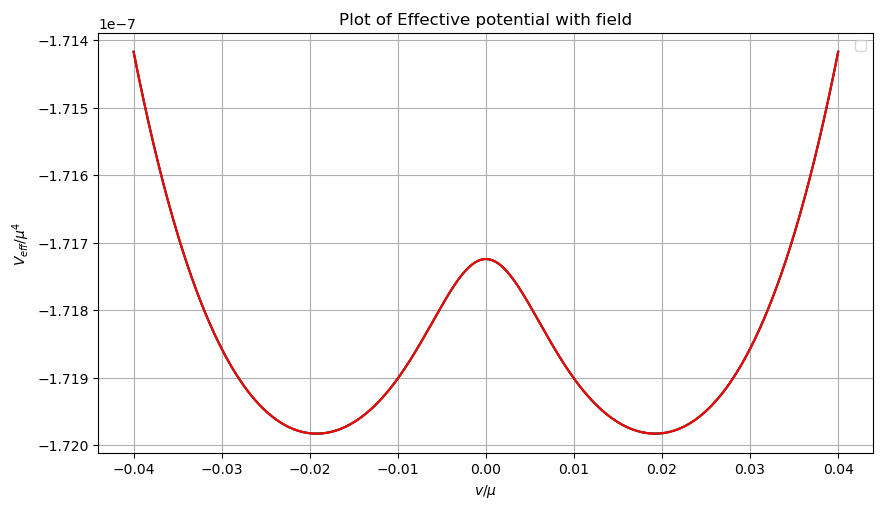}
\caption{  \it \small Variation of the effective potential, \ref{v21'}, with the background field for de Sitter spacetime, respectively for $N=1, 2, 4$, keeping all the other parameters fixed. Note that increasing $N$  reduces  the effective potential and moves it towards negative values.   }
\label{symb6}
\end{center}
\end{figure}
%
\subsection{$O(N)$ model in the broken phase}\label{S6}
Let us now come to the more interesting case of the broken phase of the $O(N)$ model, in which case only one field has a classical background while the rest $(N-1)$ are purely quantum. 
We begin by considering the same equations as in the symmetric phase \ref{v37}, \ref{v38}, \ref{v39} and \ref{v40}. 
\begin{figure}[H]
\begin{center}
\includegraphics[scale=0.90]{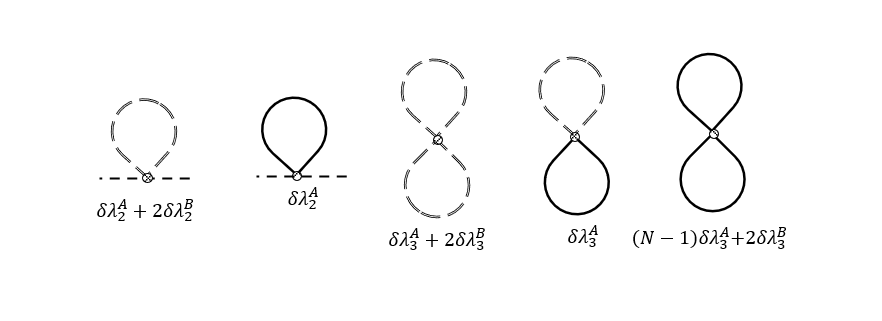}
\caption{  \it \small Diagrams pertaining to the broken phase in the Hartree approximation for the $O(N)$ model. Solid and faded dashed lines correspond respectively to $\pi$ and $\sigma$ field propagators. The  non-faded dashed lines stand for the background field.}
\label{symb103}
\end{center}
\end{figure}
We take  $v^1=v$,  $v^i=0$ for all $i= 2, 3, \cdots N-1$ and rename the quantum  fluctuations as $\varphi^1=\sigma$, $\varphi^i=\pi^i$ ($ i=2, 3, \cdots N-1$). With these variables, the 2PI effective action can be written as, 
\begin{eqnarray}
&&\Gamma_{\rm 2PI}[v,iG_{\sigma},iG_{\pi}]=-\int d^d x \sqrt{-g} \biggl[ \frac{1}{2}v\left(-\Box+m_0^2 +\delta m_1^2+(\xi+\delta \xi_1)R \right)v+\frac{(\lambda +\delta\lambda_1)v^4}{4!N} \biggr] +\frac{i}{2}\ln\det[iG_{\sigma}]\nonumber\\
&&+\frac{i(N-1)}{2}\ln\det[iG_{\pi}]  -\frac{1}{2}\mathrm{Tr}\left[\left(-\Box+m_0^2+\delta m_2^2+(\xi+\delta \xi_2)R+\frac{(3\lambda+\delta\lambda_2^A+2\delta\lambda_2^B)v^2}{6N}\right)iG_{\sigma}\right] \nonumber\\
&&-\frac{N-1}{2}\mathrm{Tr}\left[\left(-\Box+m_0^2+\delta m_2^2+(\xi+\delta \xi_2)R+\frac{(\lambda+\delta\lambda_2^A)v^2}{6N} \right)iG_{\pi}\right]-\frac{1}{4!N}\int \!d^dx \sqrt{-g} (3\lambda+\delta\lambda_3^A+2\delta\lambda_3^B)(iG_{\sigma})^2 \nonumber\\
&&-\frac{N-1}{12N}\int \! d^dx \sqrt{-g} (\lambda + \delta\lambda_3^A) iG_{\sigma}\ iG_{\pi}-\frac{N-1}{24N}\int \!d^dx \sqrt{-g} \Bigl[(N-1)(\lambda+\delta\lambda_3^A)+2(\lambda+\delta\lambda_3^B)\Bigr](iG_{\pi})^2,
\label{t9}
\end{eqnarray}
Equations of motion for $v$, $iG_{\sigma}$ and $iG_{\pi}$ derived from the above effective action reads,
\begin{equation}
\left[ -\Box+m_0^2+\delta m_1^2+(\xi+\delta \xi_1)R+\frac{(\lambda +\delta\lambda_1)v^2}{6N} +\frac{(3\lambda+\delta\lambda_2^A+2\delta\lambda_2^B)}{6N}iG_{\sigma}(x,x)+\frac{N-1}{6N}(\lambda+\delta\lambda_2^A)iG_{\pi}(x,x)\right] v(x)=0,
\label{v43}
\end{equation}
\begin{eqnarray}
&&\left[ -\Box_x+m_0^2+\delta m_2^2+(\xi+\delta \xi_2)R+\frac{(3\lambda+\delta\lambda_2^A+2\delta\lambda_2^B)v^2}{6N}+\frac{(3\lambda+\delta\lambda_3^A+2\delta\lambda_3^B)}{6N}iG_{\sigma}(x,x)\right. \nonumber\\  && \left.+\frac{N-1}{6N}(\lambda+\delta\lambda_3^A) iG_{\pi}(x,x)\right]i G_{\sigma}(x,x') =-i\delta^d(x-x'),
\label{v44}
\end{eqnarray}
and
\begin{eqnarray}
&&\left[ -\Box+m_0^2+\delta m_2^2+(\xi+\delta \xi_2)R+\frac{(\lambda+\delta\lambda_2^A)v^2}{6N} +\frac{\lambda+\delta\lambda_2^A}{6N}i G_{\sigma}(x,x)\right. \nonumber\\  && \left.+\frac{1}{6N}\bigl[ (N-1)(\lambda+\delta\lambda_2^A) +2(\lambda+\delta\lambda_3^A)\bigr] iG_{\pi}(x,x)\right]iG_{\pi}(x,x')=-i\delta^d(x-x').
\label{v45}
\end{eqnarray}
The coincidence propagators, \ref{v14}, in this case becomes 
\begin{eqnarray}
iG_{\sigma,\pi}(x,x)= m_{\sigma ,\pi\rm{dyn,eff}}^2f_d+\left(\xi -\frac16 \right)R f_d+f_{\rm{fin}}^{\sigma,\pi}
\label{v50}
\end{eqnarray}
Where $f_{\rm{fin}}^{\sigma,\pi}$ has the same form as of  \ref{v25}, with the effective dynamical   mass term replaced by $m^2_{\sigma, \pi{\rm dyn, eff}}$. From \ref{v44}, \ref{v45}, we now have the effective dynamical masses for $\sigma$ and $\pi^i$ fields,
\begin{eqnarray}
&& m_{\sigma\rm{dyn,eff}}^2 +\xi R = m_0^2+\delta m_2^2+(\xi+\delta \xi_2)R+\frac{(3\lambda+\delta\lambda_2^A+2\delta\lambda_2^B)v^2}{6N} +\frac{(3\lambda+\delta\lambda_3^A+2\delta\lambda_3^B)}{6N}\left(m_{\sigma \rm{dyn,eff}}^2f_d+\left(\xi -\frac16 \right)R f_d+f_{\rm{fin}}^{\sigma}\right) \nonumber\\&&+\frac{N-1}{6N}(\lambda+\delta\lambda_3^A)\left(m_{\pi \rm{dyn,eff}}^2f_d+\left(\xi -\frac16 \right)R f_d+f_{\rm{fin}}^{\pi}\right),
\label{t10}
\end{eqnarray}
and,
\begin{eqnarray}
&& m_{\pi\rm{dyn,eff}}^2 +\xi R = m_0^2+\delta m_2^2+(\xi+\delta \xi_2)R+\frac{(\lambda+\delta\lambda_2^A)v^2}{6N} +\frac{\lambda+\delta\lambda_2^A}{6N}\left(m_{\sigma \rm{dyn,eff}}^2f_d+\left(\xi -\frac16 \right)R f_d+f_{\rm{fin}}^{\sigma}\right)\nonumber\\&& +\frac{1}{6N}\bigl[ (N-1)(\lambda+\delta\lambda_2^A) +2(\lambda+\delta\lambda_3^A)\bigr]\left(m_{\pi \rm{dyn,eff}}^2f_d+\left(\xi -\frac16 \right)R f_d+f_{\rm{fin}}^{\pi}\right).
\label{t11}
\end{eqnarray}
From the above equation, we infer the most natural expression for the effective, dynamical mass squared
\begin{eqnarray}
&& m_{\sigma {\rm dyn, eff}}^2= m_0^2 +\frac{ \lambda v^2}{2N}+\frac{ \lambda f_{\rm{fin}}^{\sigma}}{2N} + \frac{(N-1)\lambda f_{\rm{fin}}^{\pi}}{6N},\nonumber\\ &&
m_{\pi {\rm dyn, eff}}^2= m_0^2 +\frac{ \lambda v^2}{6N}+\frac{ \lambda f_{\rm{fin}}^{\sigma}}{6N} + \frac{(N+1)\lambda f_{\rm{fin}}^{\pi}}{6N},
\label{v54}
\end{eqnarray}
Note  that self energies corresponding to different fields get mixed up in the expressions of the dynamical effective masses.
The counterterms of \ref{t10} and \ref{t11} must satisfy the following two relationships 
\begin{eqnarray}
&&\delta m_2^2+\delta \xi_2 R+\frac{(\delta\lambda_2^A+2\delta\lambda_2^B)v^2}{6N}  +\frac{(3\lambda+\delta\lambda_3^A+2\delta\lambda_3^B)}{6N}\left(m_{\sigma \rm{dyn,eff}}^2+\left(\xi -\frac16 \right)R \right)f_d+\frac{(\delta\lambda_3^A+2\delta\lambda_3^B)}{6N}f_{\rm{fin}}^{\sigma}\nonumber\\&&+\frac{N-1}{6N}(\lambda+\delta\lambda_3^A)\left(m_{\pi \rm{dyn,eff}}^2+\left(\xi -\frac16 \right)R \right)f_d+\frac{N-1}{6N}\delta\lambda_3^A f_{\rm{fin}}^{\pi} =0,
\label{t12}
\end{eqnarray}
and,
\begin{eqnarray}
&&\delta m_2^2+\delta \xi_2 R+\frac{\delta\lambda_2^A v^2}{6N}  +\frac{(\lambda+\delta\lambda_2^A)}{6N}\left(m_{\sigma \rm{dyn,eff}}^2+\left(\xi -\frac16 \right)R \right)f_d+\frac{\delta\lambda_2^A}{6N}f_{\rm{fin}}^{\sigma}\nonumber\\&&+\frac{1}{6N}\bigl[ (N-1)(\lambda+\delta\lambda_2^A) +2(\lambda+\delta\lambda_3^A)\bigr]\left(m_{\pi \rm{dyn,eff}}^2+\left(\xi -\frac16 \right)R \right)f_d+\frac{1}{6N}\bigl[ (N-1)\delta\lambda_2^A +2\delta\lambda_3^A\bigr] f_{\rm{fin}}^{\pi} =0.
\label{t13}
\end{eqnarray}
Renormalising as earlier, we have the following counterterms, 
\begin{eqnarray}
\delta\lambda^B_2=\delta\lambda^B_3 = -\frac{\lambda^2 f_d}{3N}\left(1+\frac{\lambda f_d}{3N}\right),
\label{v47}
\end{eqnarray}
\begin{eqnarray}
\delta\lambda^A_2=\delta\lambda^A_3 =-\frac{\lambda f_d}{6N}\left[1+\frac{(N+2)\lambda f_d}{6N}\right]^{-1}\left[(N+4)\lambda +\delta \lambda^B_3\right],
\label{v48}
\end{eqnarray}
\begin{eqnarray}
\delta m_2^2 =-\frac{m_0^2(N+2)\lambda f_d}{6N}\left[1+\frac{(N+2)\lambda f_d}{6N}\right]^{-1},
\label{v49}
\end{eqnarray}
and
\begin{eqnarray}
\delta \xi_2 =-\frac{\left(\xi -1/6 \right)(N+2)\lambda f_d}{6N}\left[1+\frac{(N+2)\lambda f_d}{6N}\right]^{-1}
\label{v51}
\end{eqnarray}
Likewise, the field equation \ref{v43} is self-renormalised and it should take the form,
\begin{eqnarray}
&&\left[-\Box +m_0^2+\xi R+\frac{\lambda v^2}{6N} + \frac{\lambda}{2N} f^{\sigma}_{\rm fin}+ \frac{(N-1)\lambda}{6N}f^{\pi}_{\rm fin}\right] v=0.
\label{t14}
\end{eqnarray}
We also note down further  consistent renormalisation conditions as follows
\be
\delta m_1^2= \delta m_2^2, \qquad \delta \xi_1=\delta \xi_2, \qquad \delta \lambda_1 =\delta \lambda_2^A + 2\delta \lambda_2^B.
\label{CT1'}
\ee

\noindent
In order to compute the effective action explicitly, we now note down
\begin{eqnarray}
&&\int d m_{\sigma \rm{dyn,eff}}^2 iG_{\sigma}(x,x) = \left[m_0^4+ \frac{\lambda^2 v^4}{4N^2} + \frac{\lambda^2 f^{\sigma 2}_{\rm{fin}}}{4 N^2}+\frac{(N-1)^2 \lambda^2 f^{\pi 2}_{\rm{fin}}}{36 N^2} + 2m_0^2 \left(\frac{ \lambda v^2}{2N}+\frac{ \lambda f_{\rm{fin}}^{\sigma}}{2N} + \frac{(N-1)\lambda f_{\rm{fin}}^{\pi}}{6N}\right)+\frac{\lambda^2 v^2 f^{\sigma }_{\rm{fin}}}{2N^2} \right. \nonumber\\&& \left. + \frac{(N-1)\lambda^2 v^2 f^{\pi }_{\rm{fin}}}{6N^2} +\frac{(N-1)\lambda^2 f^{\sigma }_{\rm{fin}} f^{\pi }_{\rm{fin}}}{6N^2} \right]\frac{f_d}{2} + \left(\xi -\frac16 \right)R\left(m_0^2 +\frac{ \lambda v^2}{2N}+\frac{ \lambda f_{\rm{fin}}^{\sigma}}{2N} + \frac{(N-1)\lambda f_{\rm{fin}}^{\pi}}{6N}\right)f_d +\int d m_{\sigma \rm{dyn,eff}}^2 f_{\rm{fin}}^{\sigma},
\nonumber\\&&
\label{t16}
\end{eqnarray}
\begin{eqnarray}
&&\int d m_{\pi \rm{dyn,eff}}^2 iG_{\pi}(x,x) = \left[m_0^4+ \frac{\lambda^2 v^4}{N^2} + \frac{\lambda^2 f^{\sigma 2}_{\rm{fin}}}{ N^2}+\frac{(N+1)^2 \lambda^2 f^{\pi 2}_{\rm{fin}}}{36 N^2} + 2m_0^2 \left(\frac{ \lambda v^2}{N}+\frac{ \lambda f_{\rm{fin}}^{\sigma}}{N} + \frac{(N+1)\lambda f_{\rm{fin}}^{\pi}}{6N}\right)+\frac{2\lambda^2 v^2 f^{\sigma }_{\rm{fin}}}{N^2} \right. \nonumber\\&& \left. + \frac{(N+1)\lambda^2 v^2 f^{\pi }_{\rm{fin}}}{3N^2} +\frac{(N+1)\lambda^2 f^{\sigma }_{\rm{fin}} f^{\pi }_{\rm{fin}}}{3N^2} \right]\frac{f_d}{2} + \left(\xi -\frac16 \right)R\left(m_0^2 +\frac{ \lambda v^2}{N}+\frac{ \lambda f_{\rm{fin}}^{\sigma}}{N} + \frac{(N+1)\lambda f_{\rm{fin}}^{\pi}}{6N}\right)f_d +\int d m_{\pi \rm{dyn,eff}}^2 f_{\rm{fin}}^{\pi},
\nonumber\\&&
\label{t17}
\end{eqnarray}
\begin{eqnarray}
&&(iG^{\sigma}(x,x))^2 = \left[m_0^4+ \frac{\lambda^2 v^4}{4N^2} + \frac{\lambda^2 f^{\sigma 2}_{\rm{fin}}}{4 N^2}+\frac{(N-1)^2 \lambda^2 f^{\pi 2}_{\rm{fin}}}{36 N^2} + 2m_0^2 \left(\frac{ \lambda v^2}{2N}+\frac{ \lambda f_{\rm{fin}}^{\sigma}}{2N} + \frac{(N-1)\lambda f_{\rm{fin}}^{\pi}}{6N}\right)+\frac{\lambda^2 v^2 f^{\sigma }_{\rm{fin}}}{2N^2}  + \frac{(N-1)\lambda^2 v^2 f^{\pi }_{\rm{fin}}}{6N^2} \right. \nonumber\\&& \left. +\frac{(N-1)\lambda^2 f^{\sigma }_{\rm{fin}} f^{\pi }_{\rm{fin}}}{6N^2} \right]f_d^2 +2\left(m_0^2 +\frac{ \lambda v^2}{2N}+\frac{ \lambda f_{\rm{fin}}^{\sigma}}{2N} + \frac{(N-1)\lambda f_{\rm{fin}}^{\pi}}{6N}\right)f_d \left(\left(\xi -\frac16 \right)R f_d + f^{\sigma }_{\rm{fin}} \right)+ \left(\xi -\frac16 \right)^2R^2 f_d^2 \nonumber\\&&+\left(\xi -\frac16 \right)R f_df^{\sigma }_{\rm{fin}} +(f^{\sigma}_{\rm{fin}})^2,
\label{t18}
\end{eqnarray}
\begin{eqnarray}
&&(iG^{\pi}(x,x))^2 = \left[m_0^4+ \frac{\lambda^2 v^4}{N^2} + \frac{\lambda^2 f^{\sigma 2}_{\rm{fin}}}{ N^2}+\frac{(N+1)^2 \lambda^2 f^{\pi 2}_{\rm{fin}}}{36 N^2} + 2m_0^2 \left(\frac{ \lambda v^2}{N}+\frac{ \lambda f_{\rm{fin}}^{\sigma}}{N} + \frac{(N+1)\lambda f_{\rm{fin}}^{\pi}}{6N}\right)+\frac{2\lambda^2 v^2 f^{\sigma }_{\rm{fin}}}{N^2}  + \frac{(N+1)\lambda^2 v^2 f^{\pi }_{\rm{fin}}}{3N^2}\right. \nonumber\\&& \left. +\frac{(N+1)\lambda^2 f^{\sigma }_{\rm{fin}} f^{\pi }_{\rm{fin}}}{3N^2} \right]f_d^2 +2\left(m_0^2 +\frac{ \lambda v^2}{N}+\frac{ \lambda f_{\rm{fin}}^{\sigma}}{N} + \frac{(N+1)\lambda f_{\rm{fin}}^{\pi}}{6N}\right)f_d \left(\left(\xi -\frac16 \right)R f_d + f^{\pi }_{\rm{fin}} \right)+ \left(\xi -\frac16 \right)^2R^2 f_d^2 \nonumber\\&&+\left(\xi -\frac16 \right)R f_df^{\pi }_{\rm{fin}} +(f^{\pi}_{\rm{fin}})^2,
\label{t19}
\end{eqnarray}
and
\begin{eqnarray}
&&iG^{\sigma }(x,x)iG^{\pi}(x,x) = \left(m_0^2 +\frac{ \lambda v^2}{2N}+\frac{ \lambda f_{\rm{fin}}^{\sigma}}{2N} + \frac{(N-1)\lambda f_{\rm{fin}}^{\pi}}{6N}\right)\left(m_0^2 +\frac{ \lambda v^2}{N}+\frac{ \lambda f_{\rm{fin}}^{\sigma}}{N} + \frac{(N+1)\lambda f_{\rm{fin}}^{\pi}}{6N}\right)f_d^2 \nonumber\\&&+ \left(m_0^2 +\frac{ \lambda v^2}{2N}+\frac{ \lambda f_{\rm{fin}}^{\sigma}}{2N} + \frac{(N-1)\lambda f_{\rm{fin}}^{\pi}}{6N}\right)f_{\rm{fin}}^{\pi} f_d +\left(m_0^2 +\frac{ \lambda v^2}{N}+\frac{ \lambda f_{\rm{fin}}^{\sigma}}{N} + \frac{(N+1)\lambda f_{\rm{fin}}^{\pi}}{6N}\right)f_{\rm{fin}}^{\sigma}f_d\nonumber\\&& +\left(2m_0^2 +\frac{ 3\lambda v^2}{2N}+\frac{ \lambda f_{\rm{fin}}^{\sigma}}{2N} + \frac{(N-1)\lambda f_{\rm{fin}}^{\pi}}{6N} +\frac{ \lambda f_{\rm{fin}}^{\sigma}}{N} + \frac{(N+1)\lambda f_{\rm{fin}}^{\pi}}{6N}\right)f_d^2+ \left(\xi -\frac16 \right)^2R^2 f_d^2+f_{\rm{fin}}^{\sigma}f_{\rm{fin}}^{\pi}.
\label{t20}
\end{eqnarray}

\noindent
Putting everything together now, and using the renormalisation consistency conditions, we obtain as earlier the renormalised effective action,
\begin{eqnarray}
&&\Gamma_{\rm 1PI}[v]= \int d^4x \sqrt{-g} \left[-\frac12 v\left(-\Box +m_0^2\right)v-\frac{\xi R v^2}{2}-\frac{\lambda v^4}{4!N}-\frac12 \int dm_{\sigma\rm{dyn,eff}}^2 f_{\rm{fin}}^{\sigma}-\frac{N-1}{2} \int dm_{\pi\rm{dyn,eff}}^2 f_{\rm{fin}}^{\pi} \right. \nonumber\\&& \left. +\frac{\lambda}{8N}(f_{\rm{fin}}^{\sigma })^2+\frac{(N-1)\lambda}{12N}f_{\rm{fin}}^{\sigma} f_{\rm{fin}}^{\pi}+\frac{(N-1)(N+1)\lambda}{24N}(f_{\rm{fin}}^{\pi})^2\right], 
\label{v52}
\end{eqnarray}
and accordingly, the effective potential,
\begin{eqnarray}
&&V_{\rm{eff}}(v)= \frac12 \left(m_0^2+\xi R\right)v^2+\frac{\lambda v^4}{4!N}+\frac12 \int dm_{\sigma\rm{dyn,eff}}^2 f_{\rm{fin}}^{\sigma}+\frac{N-1}{2} \int dm_{\pi\rm{dyn,eff}}^2 f_{\rm{fin}}^{\pi}  -\frac{\lambda}{8N}(f_{\rm{fin}}^{\sigma })^2-\frac{(N-1)\lambda}{12N}f_{\rm{fin}}^{\sigma} f_{\rm{fin}}^{\pi}\nonumber\\&&-\frac{(N-1)(N+1)\lambda}{24N}(f_{\rm{fin}}^{\pi})^2. 
\label{t15}
\end{eqnarray}
Setting $N=1$ in the two above equations, we reproduce the $\phi^4$ theory result, \ref{v23}, \ref{v24}. Also, in this case only the first of \ref{v54} is relevant with $f^{\pi}_{\rm fin}=0$. \\

\begin{figure}[H]
\begin{center}
\includegraphics[scale=0.195]{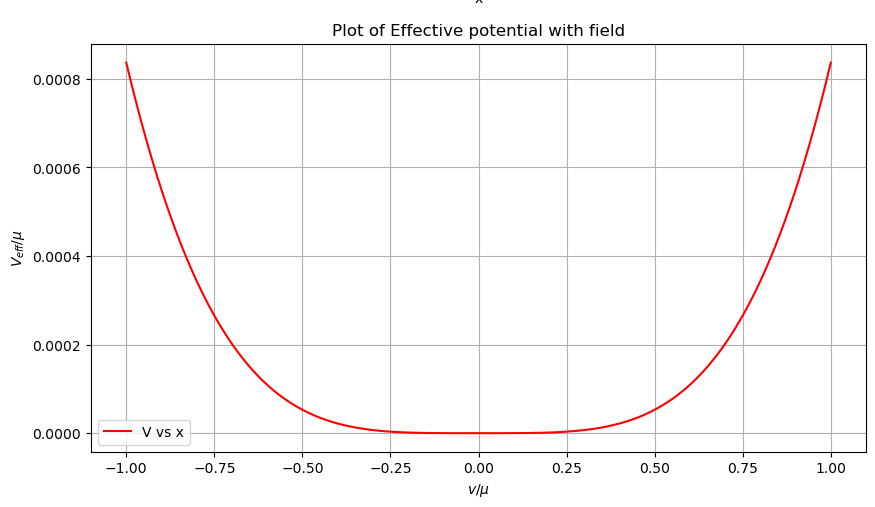}
\includegraphics[scale=0.195]{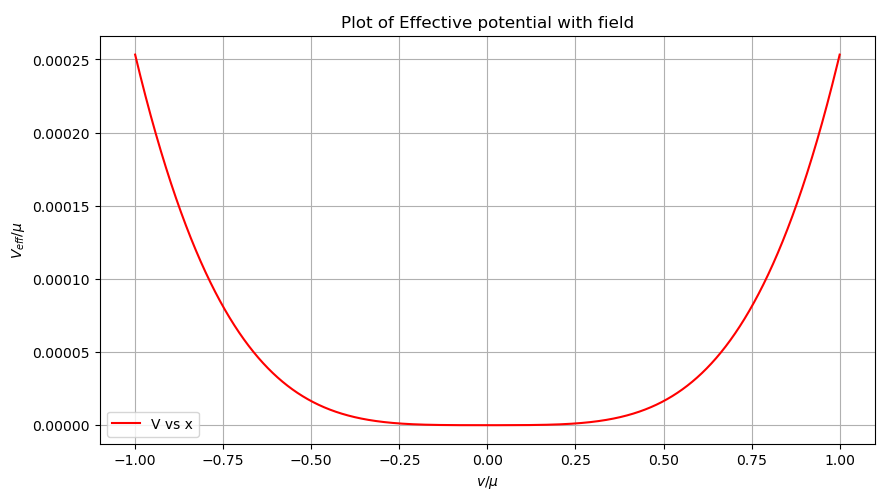}
\includegraphics[scale=0.195]{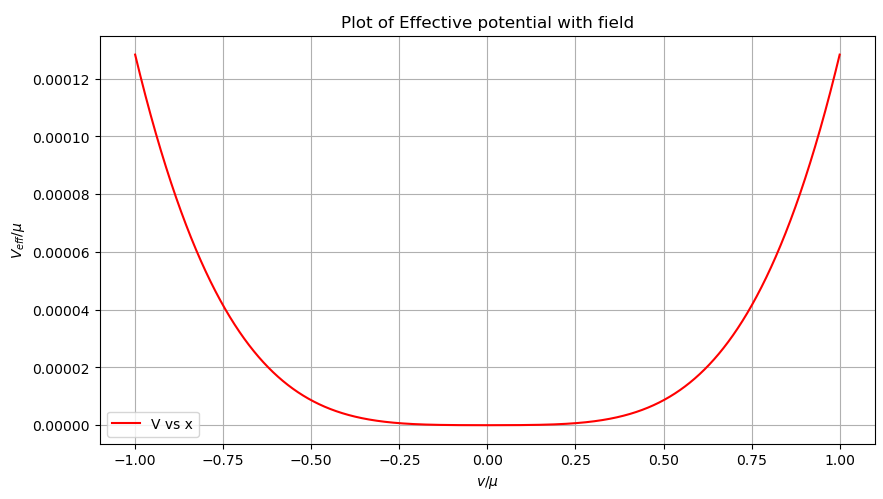}
\caption{  \it \small Variation  of the effective potential \ref{t15} with respect to the field background field for $N=3,10,20$, respectively, in the de Sitter spacetime. We have taken $m_0=10^{-5}\rm{GeV}, \xi=0,\ \lambda =0.01$, and $\Lambda = 10^{-5} \rm{GeV^2}$. Unlike the $N=1$ or the symmetric phase, there is no SSB here, in the broken phase. See main text for discussion. }
\label{symb499}
\end{center}
\end{figure}

\begin{figure}[H]
\begin{center}
\includegraphics[scale=0.22]{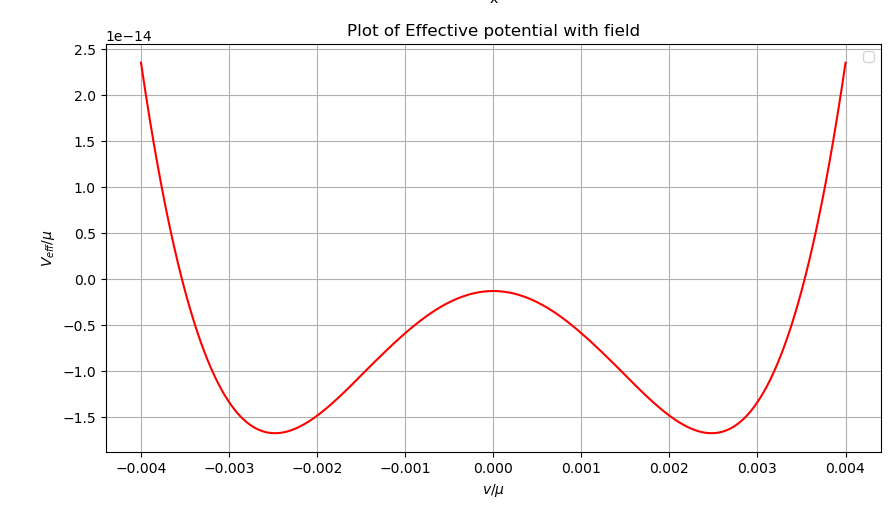}
\includegraphics[scale=0.22]{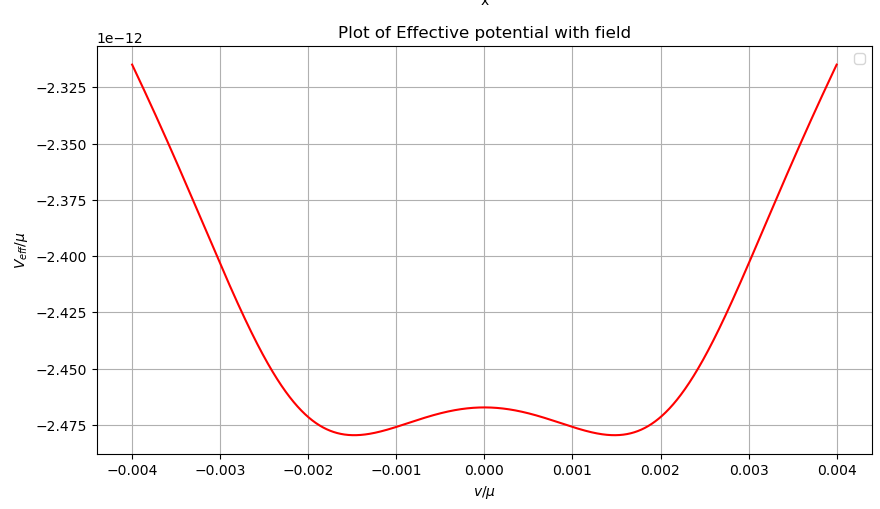}
\includegraphics[scale=0.22]{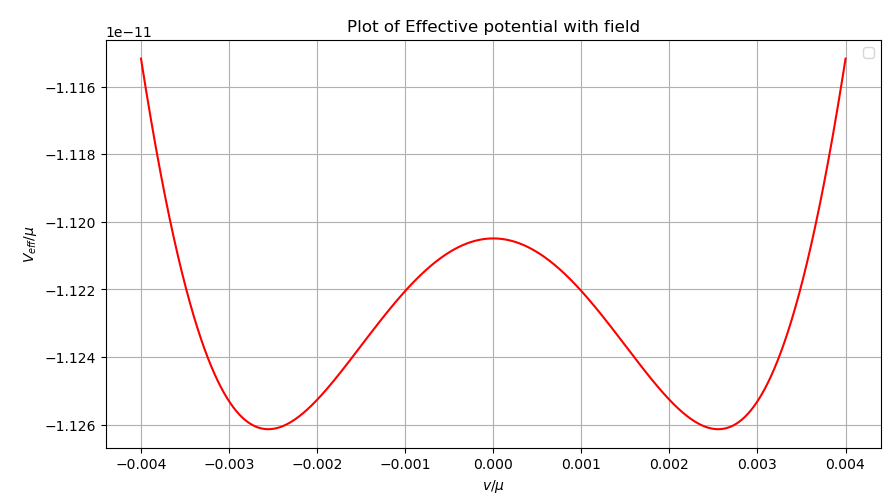}
\includegraphics[scale=0.22]{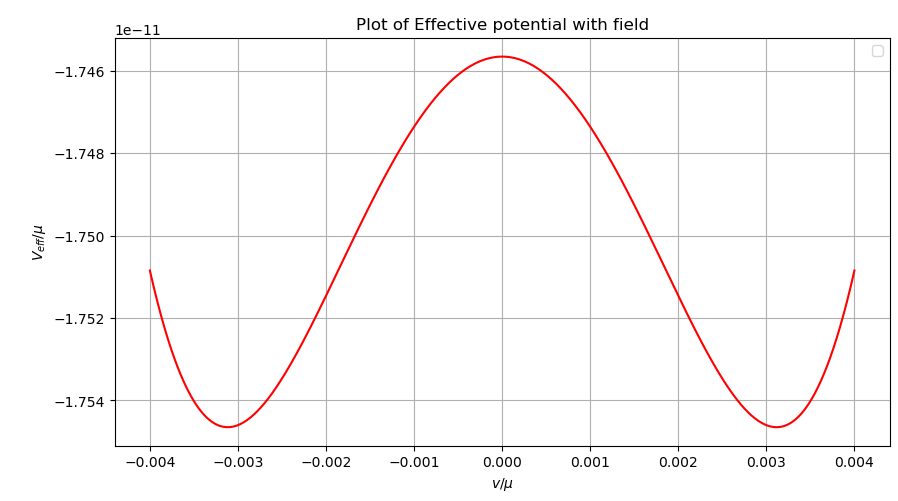}
\caption{  \it \small Plot of the effective potential \ref{t15} with  $N=2,10,15$ respectively, and $m_0^2 \ensuremath{<} 0$. We have taken $|m_0|=10^{-4}\rm{GeV}, \xi=0 , \lambda =0.01$ and  $\Lambda = 10^{-5}\rm{GeV^2}$. With increasing $N$, the height of the maximum at $v=0$ tends to increase. However, this SSB pattern will wash away if we increase the $\Lambda$ value, \ref{symb501}.}
\label{symb500}
\end{center}
\end{figure}

\begin{figure}[H]
\begin{center}
\includegraphics[scale=0.22]{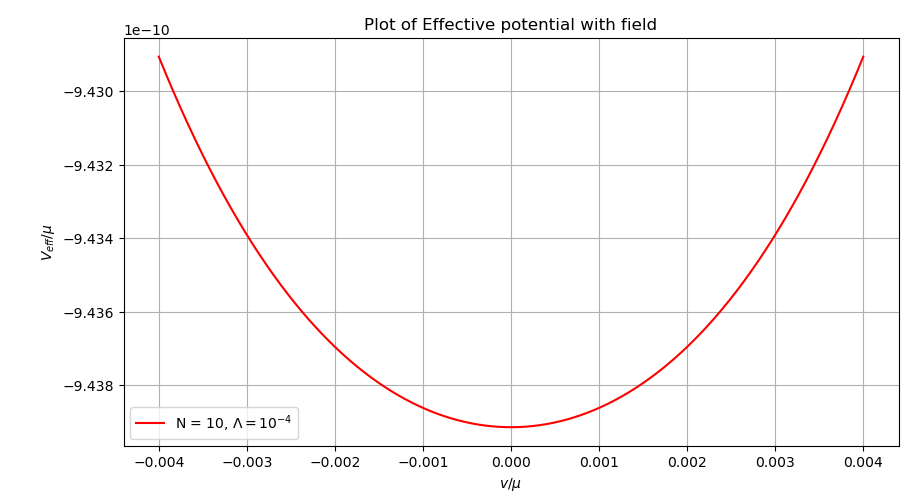}
\includegraphics[scale=0.22]{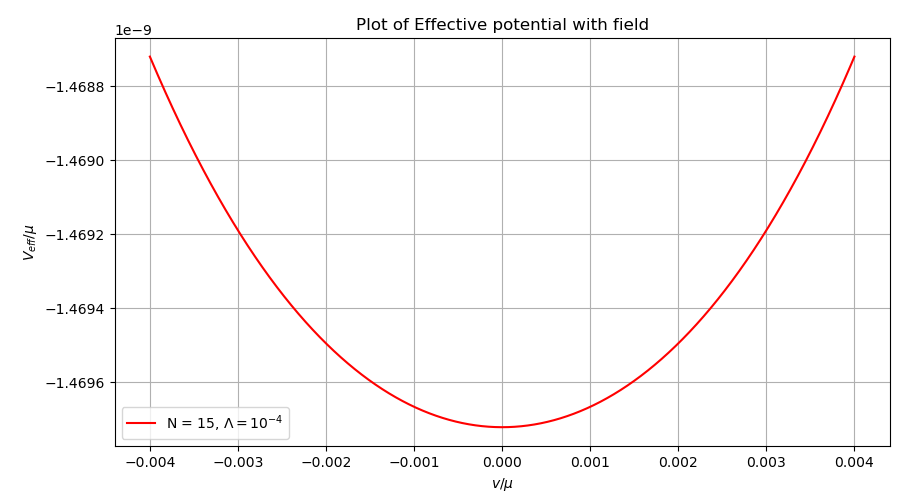}
\caption{  \it \small Restoration of broken symmetry of \ref{symb500} with increasing value of the cosmological constant, $\Lambda$. $\Lambda$ thus effectively works as temperature here.}
\label{symb501}
\end{center}
\end{figure}

\noindent
 We wish to investigate \ref{t15} for the de Sitter spacetime. In order to do this, we must compute the effective dynamical masses. However,  the immediate problem we face in \ref{v54} is the mixture of the $\sigma$ and $\pi$-fields and the non-linear equations become inseparable. Thus we need to make some approximations to find out $m^2_{\sigma, \pi {\rm dyn, eff}}$. Let us suppose, albeit crudely,  that $f^{\pi}_{\rm{fin}} $ is large compared to $f^{\sigma}_{\rm{fin}}$. We also assume that $m^2_{\pi{\rm dyn, eff}}/\Lambda \ll 1$, so that at the leading order, \ref{v54} gives
\begin{eqnarray}
&& m_{\sigma\rm{dyn,eff}}^2= m_0^2 +\frac{ \lambda v^2}{2N}+ \frac{(N-1)\lambda f_1  }{6N(4\pi)^2m_{\pi\rm{dyn,eff}}^2},\nonumber\\ &&
m_{\pi\rm{dyn,eff}}^2= m_0^2 +\frac{ \lambda v^2}{6N} + \frac{(N+1)\lambda f_1  }{6N(4\pi)^2m_{\pi\rm{dyn,eff}}^2},
\label{v58}
\end{eqnarray}
which yields,
\begin{eqnarray}
m_{\pi\rm{dyn,eff}}^2= \frac12\left(m_0^2 +\frac{ \lambda v^2}{6N}\right) \pm \frac12 \sqrt{\left(m_0^2 +\frac{ \lambda v^2}{6N}\right)^2+\frac{2(N+1)\lambda f_1}{3N(4\pi)^2}}.
\label{v59}
\end{eqnarray}
Let us first take the positive root ($m_{\pi\rm{dyn,eff}}^2{\ensuremath >}0$). Plugging it into the first of \ref{v58} gives us $m^2_{\sigma\rm{dyn,eff}}$. With these effective dynamical masses, \ref{symb499} shows the variation of the effective potential in the broken phase, \ref{t15} for different $N$ values. We see that there is no SSB, unlike the $N=1$ or the symmetric phase.   
In all the above analyses, we have taken the field rest mass squared to be positive. Let us now see what if we take it to be negative in \ref{t15} instead. It is very well known that at tree level such a potential shows SSB pattern. \ref{symb500} shows that for , there will be SSB at loop level as well. However, such pattern is washed away with increasing $\Lambda$ values, \ref{symb501}.

\begin{figure}[H]
\begin{center}
\includegraphics[scale=0.22]{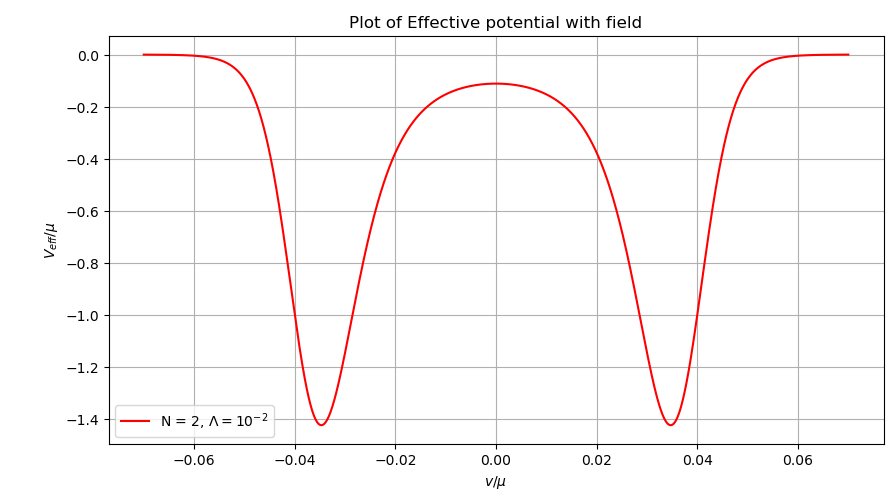}
\includegraphics[scale=0.22]{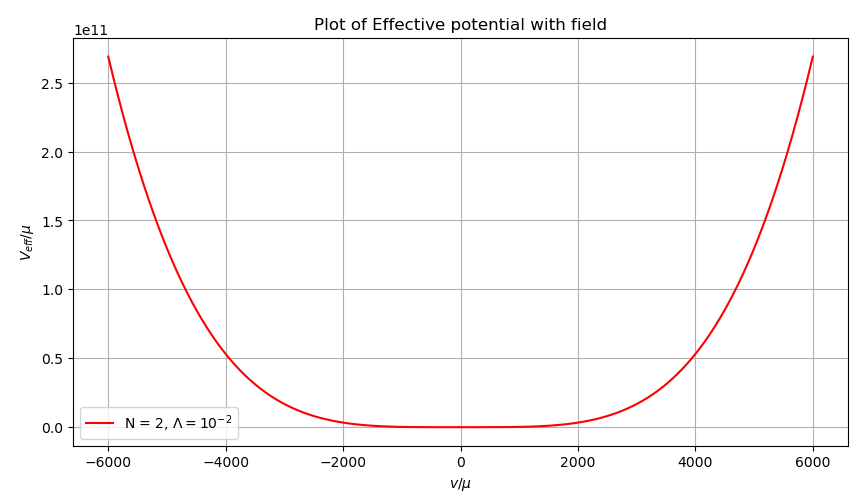}
\includegraphics[scale=0.22]{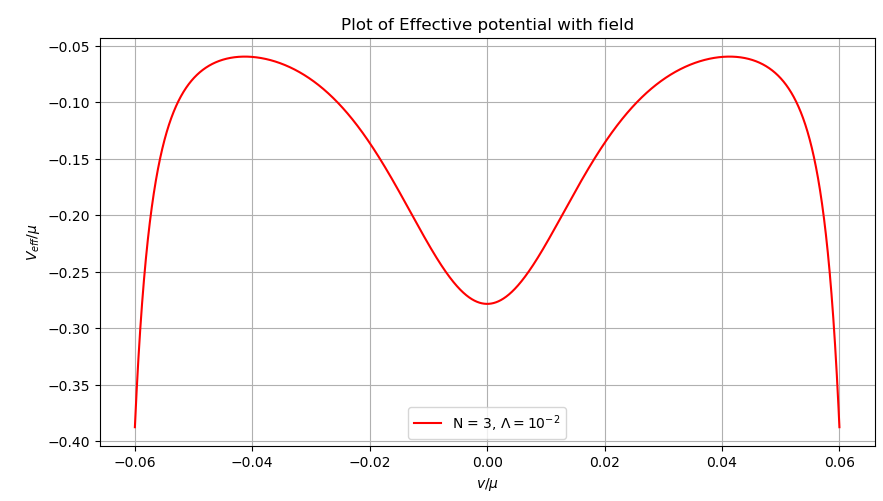}
\includegraphics[scale=0.22]{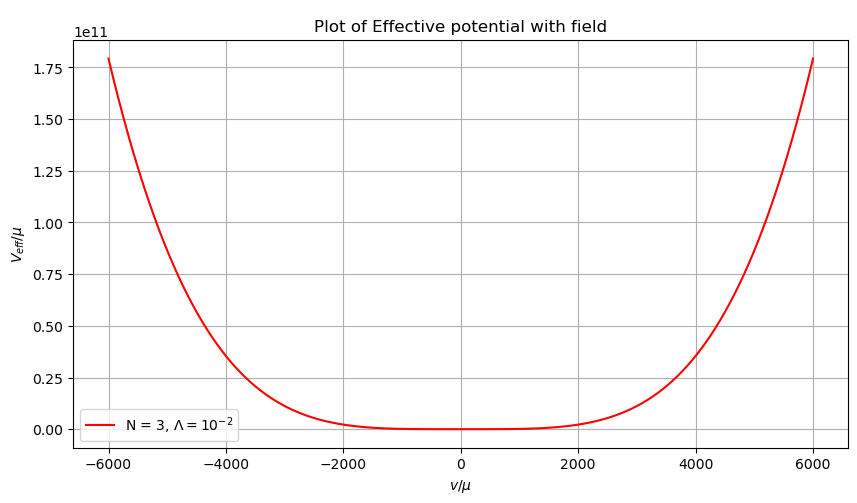}
\caption{ \it \small Plot of the effective potential \ref{t15} with  $N=2$ (first row) and $N=3$ (second row) respectively, and $m_0^2 = 0$. We have taken $ \xi=0 , \lambda =0.01$ and  $\Lambda = 0.01\rm{GeV^2}$, and a negative dynamical mass squared. We have changed the scale in each row to describe the full feature, showing both SSB and boundedness from below. Note that the pattern of the plots change drastically as we increase $N$.} 
\label{symb504}
\end{center}
\end{figure}

\begin{figure}[H]
\begin{center}
\includegraphics[scale=0.25]{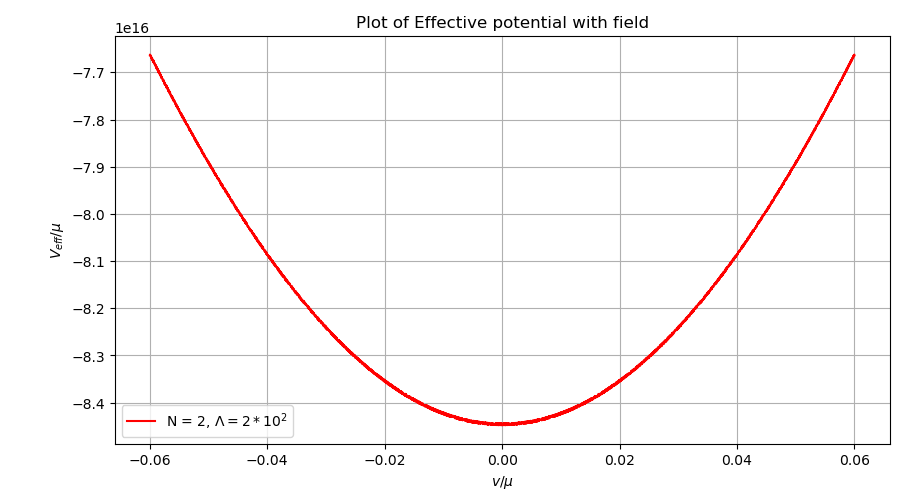}
\includegraphics[scale=0.25]{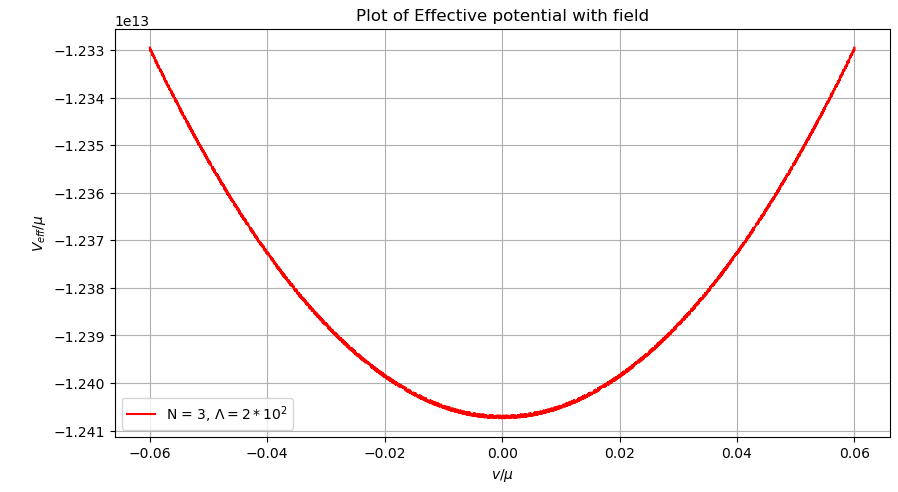}
\caption{  \it \small Plot of the effective potential \ref{t15} with  $N=2,3$ respectively, and $m_0^2 = 0$. We have taken $ \xi=0 , \lambda =0.01$ and  $\Lambda = 200\rm{GeV^2}$. Comparing with \ref{symb504}, we see that the increase in the $\Lambda$-value, has washed away the SSB pattern. The situation is similar to~\ref{symb501}.}
\label{symb505}
\end{center}
\end{figure}

\noindent
Let us now take the negative root of \ref{v59} ($m_0^2\ensuremath{<} 0$), and for simplicity set $m^2_0=0$. Note the qualitative difference with the scenario described above, for which $m_0^2\ensuremath{<} 0 $.$m_{\sigma\rm{dyn,eff}}$  is found as earlier from  the first  of \ref{v58}. With these, we have plotted \ref{t15}  for two values of $N$ in \ref{symb504}. 
Next we increase the value of $\Lambda$. We see in \ref{symb505} that as earlier the SSB pattern washes away.
%

\subsection{On the symmetry restoration at high temperature}
\begin{figure}[H]
\begin{center}
\includegraphics[scale=0.28]{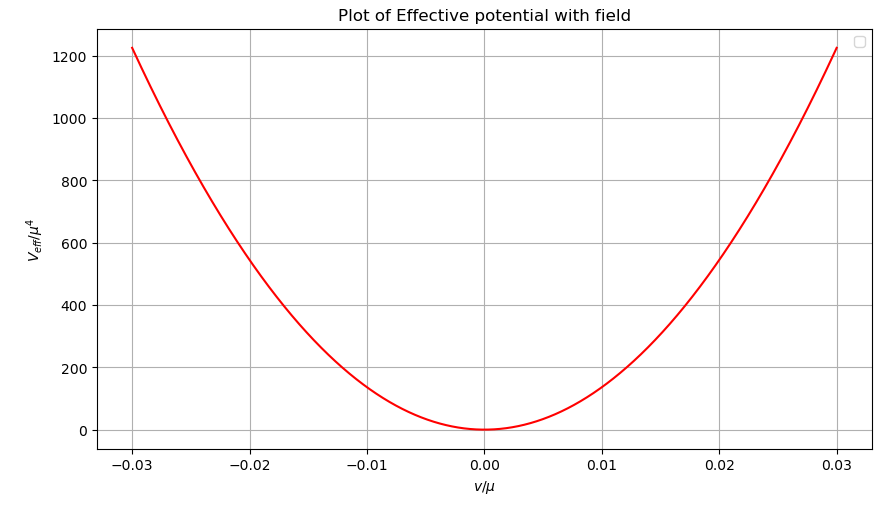}
\caption{  \it \small Plot of the finite temperature effective potential, \ref{t25}, for an $O(N)$ scalar in the symmetric phase. We have taken $m_0= 10^{-5}{\rm GeV}$, $\lambda =0.01$, $\Lambda=10^{-5}{\rm GeV}^2$, $\beta = 10^{-5} {\rm GeV}^{-1}$, $N=2$ and $\xi=0$. The symmetry breaking seen at the zero temperature \ref{symb4'}, has been restored here.}
\label{symb9}
\end{center}
\end{figure}

Finally, we wish to discuss very briefly the restoration of symmetry  for the $O(N)$ model at high temperature, as of the $N=1$ case discussed in \ref{S5}. Let us first consider the symmetric phase. 
The effective potential is formally similar to \ref{v21'}, with various finite terms replaced by temperature dependent quantities,
\begin{eqnarray}
V_{\rm{eff},\beta}(v)=N\left[\frac12(m_0^2+\xi R)v^2+\frac{\lambda v^4}{4!}-\frac{\left(1+2/N\right)\lambda f^{2}_{\beta\rm{fin}}}{24}  +\frac12 \int dm_{\beta\rm{dyn,eff}}^2 f_{\beta\rm{fin}} \right]
\label{t25}
\end{eqnarray}
Following the same procedure as of \ref{S5}, we obtain \ref{symb9} for \ref{t25} at high temperature. This  shows that there is no symmetry breaking pattern at high temperature, whereas for the same parameter values at zero temperature, there was indeed symmetry breaking (\ref{symb4'}).

\noindent
Let us now consider the effective potential at broken phase. The finite temperature effective potential  is similar to \ref{t15}, with various finite quantities temperature dependent. The two mass equations \ref{v54} look like,
\begin{eqnarray}
&& m_{\beta\sigma {\rm dyn, eff}}^{ 2}= m_0^2 +\frac{ \lambda v^2}{2N}+\frac{ \lambda f_{\beta\rm{fin}}^{\sigma}}{2N} + \frac{(N-1)\lambda f_{\beta\rm{fin}}^{\pi}}{6N},\nonumber\\ &&
m_{\beta\pi {\beta\rm dyn, eff}}^{ 2}= m_0^2 +\frac{ \lambda v^2}{6N}+\frac{ \lambda f_{\beta\rm{fin}}^{\sigma}}{6N} + \frac{(N+1)\lambda f_{\beta\rm{fin}}^{\pi}}{6N},
\label{t26}
\end{eqnarray}
where $f_{\beta\rm{fin}}^{\sigma, \pi}$ are the same as  \ref{t6}, with $S_1({\beta}),S_2({\beta})\,\, {\rm and}\,\, S_3({\beta})$  given by \ref{v32}. At very high temperature, one may ignore the curvature effect in the effective dynamical mass at the leading order,
\begin{eqnarray}
&& m_{\beta\sigma {\rm dyn, eff}}^{ 2}\simeq  m_0^2 +\frac{ \lambda v^2}{2N}+\frac{ \lambda }{48N\beta^2} + \frac{(N-1)\lambda}{144N\beta^2},\nonumber\\ &&
m_{\beta\pi {\beta\rm dyn, eff}}^{ 2}\simeq  m_0^2 +\frac{ \lambda v^2}{6N}+\frac{ \lambda }{144N\beta^2} + \frac{(N+1)\lambda }{144N\beta^2},
\label{t29}
\end{eqnarray}
We also have at this extreme high temperature,
$$f^{\pi}_{\beta\rm{fin}}=f^{\sigma}_{\beta\rm{fin}}\simeq \frac{1}{24\beta^2}$$
Thus at extreme high temperature the effective potential is curvature independent and hence there will be no question of SSB. At not so high temperature, \ref{t26} can be attempted to be solved as of \ref{v58}. However, as we increasing $T$, it will coincide with the above extreme high $T$ limit.

\section{Conclusion}\label{concl}
In this work we have investigated spontaneous symmetry breaking or its restoration driven by background spacetime curvature, via the non-perturbative 2PI formalism in two loop Hartree or local approximation, for a scalar field theory with quartic self interaction. We have used the standard Schwinger-DeWitt expansion of the Feynman propagator in curved spacetime~\cite{Dewitt}. Such expansion, based upon the normal coordinate and local Lorentz invariance in a small neighbourhood of  a given spacetime point, is meant to capture the curvature effect on local or ultraviolet physical phenomenon. It cannot capture the deep infrared effects occurring at large scales, such as the secular growth in de Sitter, e.g.~\cite{Bhattacharya:2025jtp} (also references therein) or any non-local amplitude.  Nevertheless, the present formalism may be very relevant for example, when one wishes to probe initial or transient quantum processes that can occur in a cosmological or black hole  background, or in the instances for which the field has a compact support.

Recently it was shown  by taking the Schwinger-DeWitt expansion up to the  quadratic  order in curvature that spontaneous symmetry breaking (SSB) is  possible for a scalar field theory with a quartic self interaction, positive rest mass squared {\it and} a positive non-minimal coupling parameter,~\cite{Nath:2024doz}. SSB  in a flat spacetime with positive rest mass squared is not possible. Also, such SSB remains absent at the linear order in the curvature.  Since the analysis of~\cite{Nath:2024doz} is perturbative (both in coupling as well as in curvature), it is legitimate to ask, what happens in particular, if we go on increasing the order of the curvature expansion? Does the SSB pattern wash away or remain there? Or, is there any significant quantitative changes? This was the chief motivation to  address this problem non-perturbatively.

In order to do this, we resorted to the non-perturbative 2PI technique in the two loop, local or Hartree approximation in this paper. The 2PI formalism  not only resums the self energy, but also as we have discussed, some curvature terms through it. By taking the explicit example of the de Sitter spacetime, we have confirmed that such curvature driven SSB is indeed possible for $\phi^4$ theory, \ref{S3}. We have discussed extension of this result for $O(N)$ model, \ref{ON0}. In the broken phase in particular, we have shown that for $N\ensuremath{>}1$, there is no SSB if we take the dynamical mass corresponding to the $\pi$-field to be positive. However, in this phase SSB can be present with negative rest mass squared, and the SSB pattern becomes more intense with increasing $N$. This pattern has been shown to wash away with increasing value of the cosmological constant, \ref{S6}. We  have also shown that if we take the dynamical mass squared  corresponding to the $\pi$-field to be negative, there can be SSB, which again washes away with increasing $\Lambda$.  Restoration of the broken symmetry at high temperature has also been briefly discussed. As an outcome of the resummation, we have found SSB in this with vanishing non-minimal coupling in all cases, in contrast to that of the perturbative result of \cite{Nath:2024doz}. To the best of our knowledge and understanding, the results found in this paper are interesting in their own right. 

To summarise, the present paper, along with~\cite{Nath:2024doz} shows that curvature driven SSB may be possible in short scale, ultraviolet physical processes. Now,  in order to understand  whether it has any serious phenomenological consequence or not, we must look into more realistic models, such as those containing non-Abelian gauge fields and scalars and do a thorough phenomenological analysis. Furthermore, can the mass of a particle, after gaining contributions from the spacetime curvature through symmetry breaking, decay and release energy to affect the reheating procedure?  We hope to return to these interesting  issues in our future publications.

\section*{Acknowledgement} The work of VN is supported by the Junior Research Fellowship of the University Grants Comission, Govt. of India (Ref. no.~211610068936). The authors would like to thank S.~Gangopadhyay for some useful comments.



\end{document}